\documentclass[preprint,12pt,authoryear]{elsarticle}

\usepackage{amssymb}
\usepackage{makecell} 
\usepackage{geometry,xcolor,tabularx,nccmath}
\usepackage{booktabs,caption,subcaption,mathtools,multirow,graphicx}

\usepackage{amsmath,amsthm,enumitem}

\usepackage[ruled,vlined,linesnumbered]{algorithm2e}

\usepackage{lineno}
\usepackage{tikz}
\usepackage{moreverb,url}
\usepackage{rotating}
\usepackage{pgfplots}

\usepackage[colorlinks=true,bookmarksopen=true,bookmarksnumbered=true,citecolor=red,urlcolor=red]{hyperref}
\usepackage[nameinlink,capitalise]{cleveref}

\usepackage{threeparttable}

\journal{Transportation Research Part C: Emerging Technologies}

\begin{document}

\begin{frontmatter}

\title{Modeling of Mobility and Energy Policies in an Agent-Based Framework: Case Studies for Chicago Region in 2050}

\author[1]{Md Rakibul Alam}
\ead{malam@anl.gov}

\author[1]{Omer Verbas\corref{cor1}}
\ead{omer@anl.gov}

\author[1,2]{Taner Cokyasar}
\ead{tcokyasar@anl.gov}
\ead{tcokyasar@tamu.edu}

\author[1,3]{Hadi Bhidya}
\ead{hbhidya@anl.gov}

\author[4]{Jesse Altman}
\ead{jaltman@cmap.illinois.gov}

\author[4]{Nora Beck}
\ead{nbeck@cmap.illinois.gov}

\author[1]{Joshua Auld}
\ead{jauld@anl.gov}

\author[1]{Pedro Veiga de Camargo}
\ead{pveigadecamargo@anl.gov}

\author[1]{Jamie Cook}
\ead{james.cook@anl.gov}

\author[1]{Felipe de Souza}
\ead{fdesouza@anl.gov}

\author[1]{Gopindra Nair}
\ead{gnair@anl.gov}

\author[1]{Hyunseop Uhm}
\ead{huhm@anl.gov}

\author[1]{Jan Zill}
\ead{jzill@anl.gov}

\cortext[cor1]{Corresponding author}

\address[1]{Argonne National Laboratory, 9700 S. Cass Ave, Lemont, IL, 60439, USA}
\address[2]{Texas A\&M University, Fermier Hall, 106 Ross St., College Station, TX, 77843, USA}
\address[3]{The University of Tennessee, Knoxville, 1331 Circle Park Dr, Knoxville, TN, 37916, USA}
\address[4]{Chicago Metro. Agency for Planning, 433 W Van Buren St, Chicago, IL 60607, USA}

\begin{abstract}
Metropolitan regions are simultaneously pursuing several interventions to improve mobility, accessibility, and energy efficiency, necessitating integrated tools to understand how these policies interact to affect travel behavior, energy use, and infrastructure needs. This paper evaluates the combined impacts of electrification, freight demand management, road pricing, parking reform, and transit expansion on the Chicago metropolitan transportation system in 2050, using a business-as-usual (BAU) scenario as the baseline. We employ POLARIS, a large-scale agent-based modeling framework calibrated to 2019 conditions, to simulate nine policy scenarios for the seven-county northeastern Illinois region. The framework co-simulates activity-based passenger demand, endogenous freight generation, multimodal traffic assignment, and transit operations, with charging infrastructure and freight operations optimized for each case. Our findings reveal that under the high electrification scenario, total fuel mass declines by 68\% while total charging energy increases by approximately 4--8$\times$ from BAU, resulting in a peak power demand near 4 GW concentrated in the urban core. Furthermore, freight management policies reduce freight VMT by increasing trip frequency but shortening distances, smart road pricing most effectively reduces auto VMT, and transit expansion boosts ridership by 18\% relative to BAU. By presenting the first integrated, agent-based scenario framework for Chicago that jointly evaluates these interventions, this study provides actionable insights for regional transportation planning, grid infrastructure investment, and emissions reduction, highlighting the value of targeted charger upgrades and coordinated policy bundles.
\end{abstract}


\end{frontmatter}

\section{Introduction}\label{sec:intro}
Metropolitan transportation systems worldwide are undergoing several transitions such as the electrification of passenger and freight vehicles, the restructuring of freight logistics networks, and the introduction of road-pricing mechanisms. These transitions reshape travel behavior, infrastructure planning, and policy evaluation in ways that conventional travel-demand models cannot comprehensively capture due to differences in model architecture and inherent limitations of independent studies \cite{Acosta2026ANT}. For planning agencies such as the Chicago Metropolitan Agency for Planning (CMAP), which serves as the federally designated metropolitan planning organization (MPO) for the seven-county northeastern Illinois region, the challenge is to evaluate electrification, freight logistics, road pricing, etc. as coherent bundles of strategies capable of reducing energy consumption, improving network efficiency, and maintaining fiscal viability through the 2050 planning horizon \cite{rtp_2025}.

This study uses planning and operations language for agent-based regional integrated simulation (POLARIS), a large-scale agent-based demand modeling framework developed by Argonne National Laboratory \cite{auld_polaris_2016}. POLARIS co-simulates activity-based passenger demand, endogenous freight generation, multimodal traffic assignment, and transit operations. The Chicago region model was calibrated and validated against transit boardings, mode shares, departure-time distributions, and trip-length profiles to establish behavioral fidelity before projecting any future year. A foundational study by \citet{Acosta2026ANT} then applied this calibrated platform to a set of 2050 scenarios. Compared \citet{Acosta2026ANT}, this study has a different business-as-usual transit network based on updated CMAP input. Moreover, the transit expansion scenarios and the pricing scenarios also have different designs in both studies. Finally, a lot of freight-related interventions have been added to this study. All these scenarios are evaluated against the 2050 business-as-usual (BAU) as the primary baseline.



This paper makes three primary contributions:
\begin{enumerate}[leftmargin=*]
\item We are using a fully integrated framework POLARIS \cite{auld_polaris_2016} that captures all the endogeneities between traveler and firm behavior; the timing, destination, mode, and routing choices; and multimodal traffic.
\item We present a comprehensive 2050 scenario framework for the Chicago region that spans vehicle electrification, freight operations, pricing and transit expansion.
\item We quantify system-level outcomes with the metrics of vehicle-miles traveled (VMT), passenger-miles traveled (PMT), vehicle-hours traveled (VHT), passenger-hours traveled (PHT), average speed, trip duration, monetary and time cost, and energy consumption across scenarios, providing actionable insights for regional planners and policymakers.
\end{enumerate}

All scenarios are evaluated against the BAU as the primary baseline.The results are intended to inform two key efforts: CMAP's next regional transportation plan cycle and Commonwealth Edison's (ComEd's) infrastructure investment priorities as Illinois' largest electric utility.

The remainder of the paper is organized as follows. \Cref{sec:lit} section reviews the relevant literature on regional freight modeling, electrification, road pricing, and transit planning. \Cref{sec:method} section describes the POLARIS modeling framework, the fused 2050 network, and scenario design. \Cref{sec:results} section presents results for each scenario in terms of the relevant metrics. Finally, \Cref{sec:conclusion} section summarizes the results, policy implications, and directions for future work.

\section{Literature Review}\label{sec:lit}
We begin with an overview of recent literature on freight modeling in agent-based simulation frameworks, then turn to electrification, road pricing and transit.

Regional freight modeling has historically relied on fixed origin-destination matrices derived from commodity flow surveys, an exogenous source of freight vehicle demand. This decoupled approach fails to capture the feedback between freight vehicle volumes, network congestion, mode shift among freight modes, etc. which are essential for evaluating integrated transportation policies. Agent-based models (ABMs) offer a more behaviorally robust alternative for endogenous freight trips by synthesizing firms, supply-chain relationships, and delivery tours. 

Initially \citet{roorda2010conceptual} laid the foundation for modern agent-based freight modeling by organizing decisions into operational levels across commodity, logistics, traffic, and infrastructure markets. This layered structure, which connects agents through decision chains, became the basis for later models. SimMobility Freight \citep{sakai2022household} was built on this idea by combining long-, medium-, and short-term freight decisions within the broader SimMobility platform. \citet{stinson_introducing_2022} introduced CRISTAL (Collaborative, Informed, Strategic Trade Agents with Logistics), which could generate a synthetic landscape of firms and supply-chain relationships from which freight trips could emerge endogenously. This approach captured shipper-carrier selection, delivery tour routing, and congestion interactions that fixed-demand matrices cannot represent. \citet{zuniga-garcia_freight_2023} extended this line of work by developing a freight asset choice model which could predict fleet ownership (medium- and heavy-duty (MDHD) trucks) and distribution center control from more than 11-million US establishments. Applying to the Chicago metropolitan area in POLARIS, the model found that roughly 8\% of firms own HD trucks, 20\% own MD trucks, and 3\% operate distribution centers (DCs), with fleet density concentrated in known industrial corridors. \citet{cokyasar_optimization_2023} further developed route-clustering optimization within POLARIS to reduce truck VMT through shipment consolidation. Compared with other international ABM freight platforms such as SimMobility Freight \cite{sakai2020simmobility} and MASS-GT \cite{de2018empirical}, POLARIS-CRISTAL uniquely co-simulates passenger and freight demand on a shared congested network, capturing spillover effects invisible to decoupled models. The present paper exploits this capability for freight scenarios.

In recent years, electrification research has increasingly focused on the systemic consequences for road networks and electricity grids. Prior work using POLARIS for the Chicago region demonstrated that total regional charging energy consumption increases by up to 650\% under a high electrification scenario relative to the BAU \cite{Acosta2026ANT}. Temporal charging profiles vary substantially by user groups. \citet{acosta-sequeda_interdependencies_2025} demonstrated the interdependencies between electricity consumption and transit demand in urban settings, underscoring that electrification does not affect the transportation and energy sectors in isolation.

In recent years, road pricing has received sustained academic and policy attention as a mechanism for simultaneously managing travel demand, reducing emissions, and recovering revenues, when fuel-tax is lost under electrification. Empirical studies showed that cordon and distance-based road pricing hurt lower-income people more if nothing is done to help. But fairness could be improved significantly by giving income-based discounts and by using the money collected to support the people affected \cite{venezia2023equity}. A Chicago Fed Economic Perspectives article reviews how EVs erode motor fuel tax bases and concluded that a flexible road user charge tied to mileage, vehicle weight, and possibly income or emissions is the most structurally robust replacement \cite{laguardia2023electric}. The present paper extends this analysis to three additional pricing instruments: 1) a flat motor fuel tax (MFT) replacement rate applied uniformly across times of day and vehicle types; 2) a smart road user charge (RUC) that exempts low-income households (at or below 80\% of regional median income) from variable-rate pricing; and 3) expressway access fees on currently un-tolled Chicago-loop-oriented corridors. 

Historically, public transit remains the primary alternative to single-occupancy vehicle travel in the Chicago region. Transit performance is a central concern for CMAP's regional planning \cite{CMAP2018ONTO2050}. \citet{verbas_exploring_2015} established foundational methods for optimal frequency allocation across transit networks. They demonstrated that service supply is the dominant lever for ridership response. \citet {ng_redesigning_2024} extended this to multimodal network redesign under shared autonomous mobility, showing that the interaction between transit and emerging services requires integrated optimization. The POLARIS-based transit modeling framework co-simulates transit assignment with passenger and freight demand in a fully integrated environment. Prior results for the Chicago region showed that the transit expansion\~(+34\% scheduled trips in 2050) produces a 21.5\% increase in boardings relative to 2019, an elasticity of approximately 0.64, and that a further transit expansion package\~(+62\% trips over BAU) adds another 27\% in boardings \cite{Acosta2026ANT}.

\section{Methodology}
\label{sec:method}
This case study uses POLARIS \cite{auld_polaris_2016}, which integrates activity-based travel demand, freight modeling, and multimodal traffic and transit assignment to simulate person and freight movements over a 24-hour period. It is particularly well-suited for evaluating transportation systems under changing conditions, as it captures the interactions between agents, infrastructure, and services in a unified simulation environment.

\subsection{The POLARIS modeling framework}

Travel demand in POLARIS is generated using an activity-based approach. A synthetic population of individuals and households is constructed using demographic data from the US Census, Public Use Microdata Areas (PUMAs), and the American Community Survey. Long-term choices such as home, school, and work locations are assigned following established procedures \cite{auld_efficient_2010}. Based on personal and household characteristics, each individual generates a sequence of daily activities, with start times and durations estimated using hazard-based models \cite{auld_dynamic_2011}. Activity location is chosen using a multinomial logit model, while mode choice is modeled using a nested logit formulation \cite{auld_activity_2012}. In addition to timing, location, and mode, POLARIS also models travel party decisions. A schedule conflict resolution model ensures feasibility within and across household members \cite{auld_framework_2009}. This approach captures the temporal and spatial structure of daily travel and enables the model to respond to changes in travel conditions across multiple dimensions.

POLARIS includes a freight and commercial vehicle component that operates in conjunction with the passenger travel model. Using the CRISTAL model \cite{stinson_introducing_2022}, a synthetic landscape of firms and establishments is generated. These establishments produce truck and delivery trips, including e-commerce and service-related movements \cite{cokyasar_optimization_2023,davatgari_electric_2024}. The freight module reflects the interdependencies between different types of travel: for example, increased truck volumes can degrade transit reliability or contribute to roadway congestion, indirectly affecting passenger behavior. These interactions are explicitly captured in the simulation. Once the demand side is established, trips are routed using a time-dependent intermodal algorithm \cite{verbas_time-dependent_2018}. The routing procedure accounts for various modes: private auto, transit, walk, bike, Transportation Network Companies (TNCs), and freight within an integrated multimodal network. Intermodal combinations (e.g., walk-to-transit, drive-to-transit, TNC-transit) are also supported.

Traffic is simulated using a mesoscopic, multi-class traffic flow model in Lagrangian coordinates \cite{de_souza_polaris-lc_2024}, where each vehicle class (e.g., cars, trucks, buses) has distinct speed-spacing relationships. This improves on previous average-speed models by capturing class-specific congestion dynamics. Transit vehicles interact with the same road network as private vehicles, while rail services are modeled based on GTFS schedules and operate on noncongestable links. Passenger-transit interactions are also simulated, including boarding, seating, standing, and alighting. TNC services are modeled in detail, including pooled rides, corner-to-corner pickups, first/last-mile integration with transit, and more centralized or automated service configurations \cite{gurumurthy_integrating_2020, zuniga-garcia_integrating_2022}. These services operate dynamically, with supply and demand matched during the simulation.

A key feature of POLARIS is its feedback structure. The simulation operates iteratively, allowing agents to adapt based on experienced travel conditions. If a trip takes significantly longer than expected, agents may revise their routing or mode choices in subsequent iterations. Late arrivals to activities may lead to schedule adjustments, shortened activity durations, or cancellations. These behavioral responses are embedded within a dynamic network equilibrium process, where route and mode choices reflect evolving network states. This feedback architecture allows the system to capture emerging behaviors and adaptive strategies.

POLARIS has been applied in several large-scale regional studies, including technology assessments involving automation and shared mobility \cite{rousseau_smart_2020}, ride-hail simulation and validation in the Chicago region, and transit optimization efforts \cite{verbas_impact_2024, verbas_modeling_2024}. Its design enables the modeling of multimodal systems with detailed behavioral and operational mechanisms, making it well-suited for scenario analysis across a range of policy, infrastructure, and technology contexts.

\subsection{Network}
\begin{figure}[!htbp]
    \centering
    \includegraphics[width=1\linewidth]{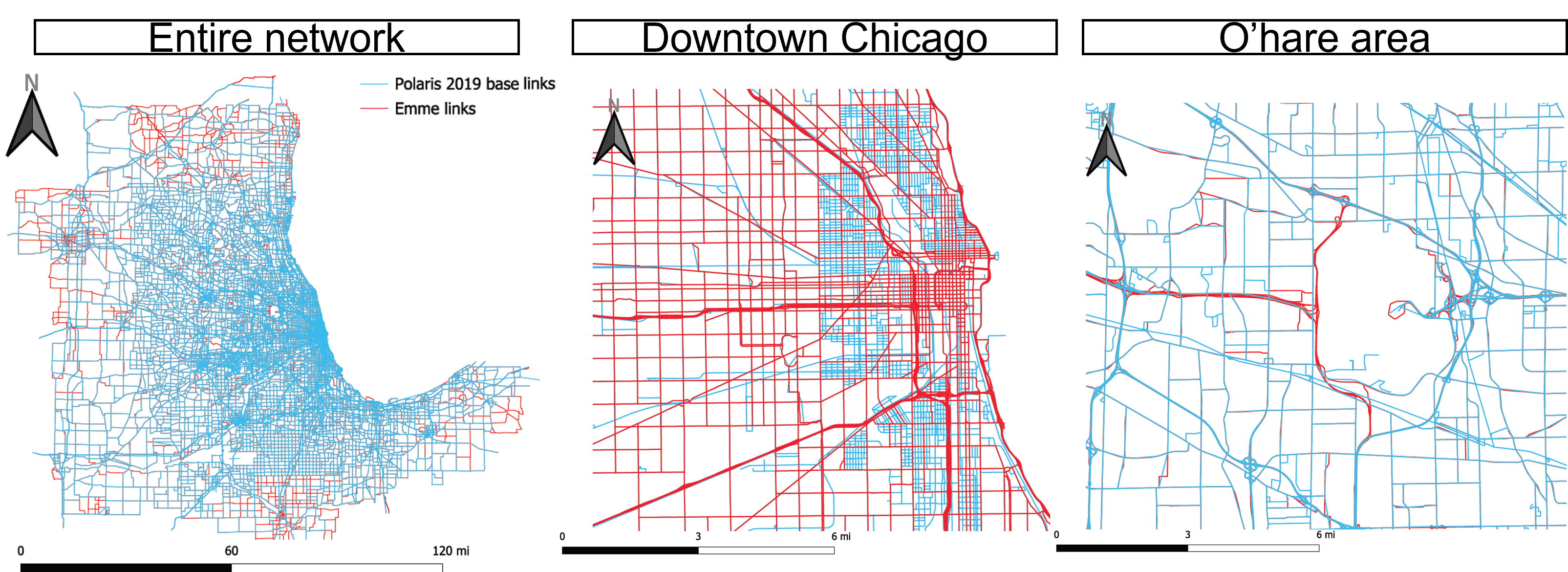}
    \caption{Chicago area Emme and POLARIS networks.}
    \label{fig:networks}
\end{figure}

A key component of this study involved developing an accurate and comprehensive network representation for the Chicago region in 2050. We began with two existing sources: a 2019 POLARIS network previously used in regional studies, and CMAP’s forecast Emme (a transportation planning software) network for 2050. While the POLARIS network offered greater granularity within the city of Chicago, the Emme network provided higher resolution in the surrounding suburban areas, as shown in \Cref{fig:networks}. To leverage the strengths of both, we created a fused network that integrated the detailed urban core of POLARIS with the suburban specificity of Emme. This fusion was guided not only by technical alignment, but also by the need to preserve elements relevant to ongoing infrastructure planning and stakeholder engagement. The process was semi-automated: while most links and nodes were merged algorithmically, conflicts,especially those affecting connectivity, were addressed through manual edits and validations. The result was a hybrid 2050 network that maximized spatial coverage, density, and relevance to both current infrastructure and future project scenarios.

\subsection{Population Projection}
\begin{figure}[!htbp]
    \centering
    \begin{subfigure}[c]{0.51\linewidth}
        \centering
        \includegraphics[width=\linewidth]{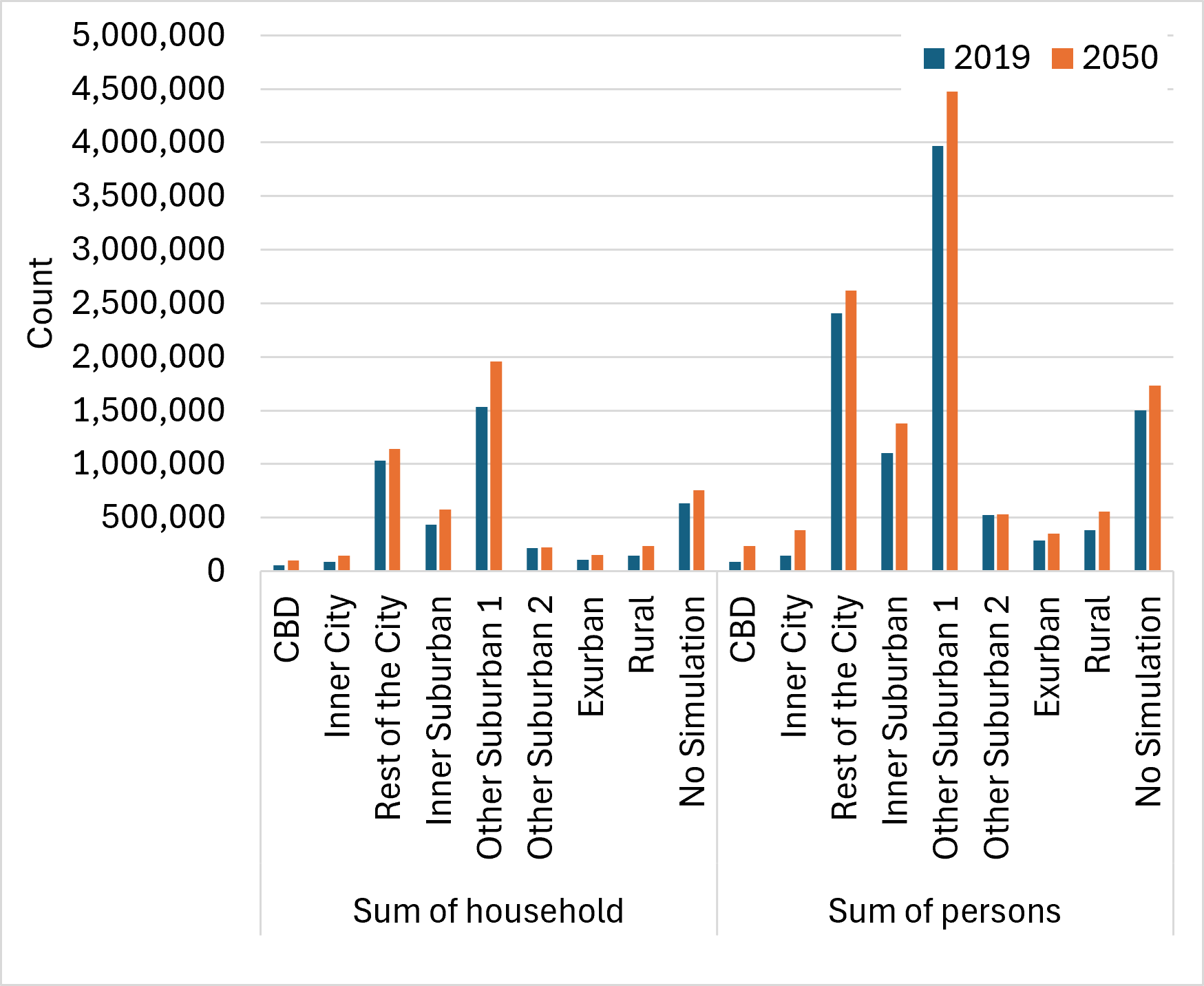}
        \caption{Population changes}
        \label{fig:pop_map}
    \end{subfigure}
    \hfill
    \begin{subfigure}[c]{0.47\linewidth}
        \centering
        \includegraphics[width=\linewidth]{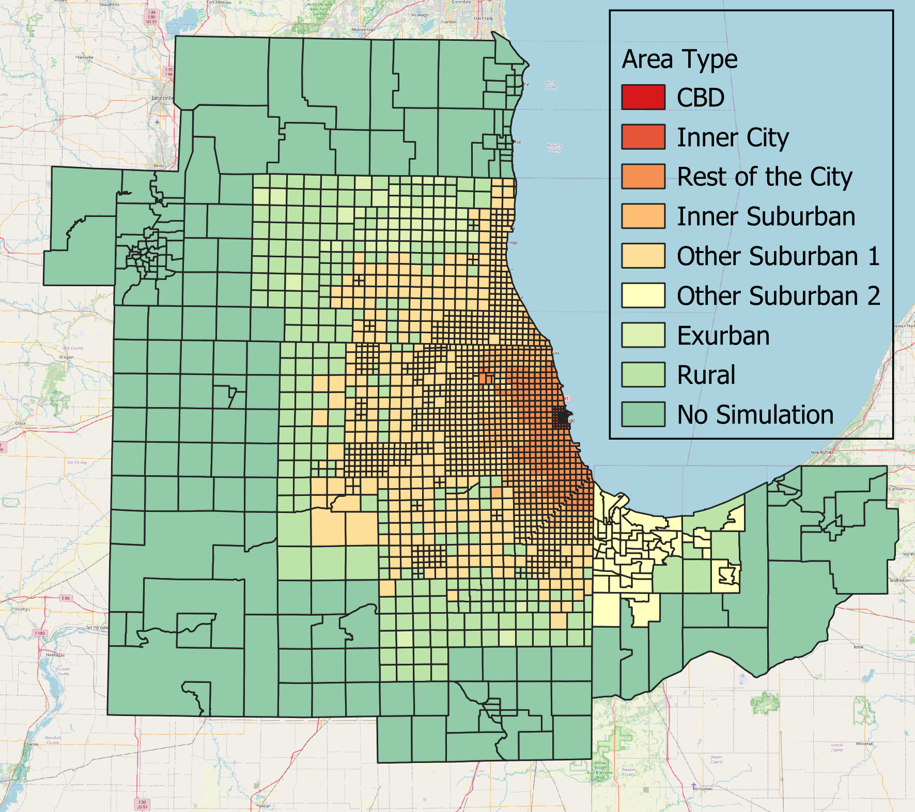}
        \caption{Chicago region area types}
        \label{fig:pop_charts}
    \end{subfigure}
    \caption{Population projections for the Chicago region under the 2050 scenario.}
    \label{fig:pop_figures}
\end{figure}

Population projections for the Chicago region reveal a considerable shift \cite{CMAP2018ONTO2050}. Under the 2050 scenario, the total household number increases from 4.3 million to 5.3 million, and the total population increases from 10.4 million to 12.4 million, compared to 2019 (\Cref{fig:pop_map}). Growth is distributed unevenly across area types (\Cref{fig:pop_charts}). The Central Business District (CBD) and Inner City experience the highest relative growth rates (roughly 2.65 times and 1.7 times increases in persons and households, respectively).

\subsection{Scenario Descriptions}
To systematically evaluate the long-term impacts of emerging transportation policies and technologies in the Chicago metropolitan region, nine modeling scenarios were constructed and organized into six thematic groups. All scenarios were executed within the POLARIS agent-based travel demand framework. The BAU serves as the primary baseline against which all policy scenarios are compared.

\begin{figure}[!htbp]
    \centering
    \includegraphics[width=0.7\linewidth]{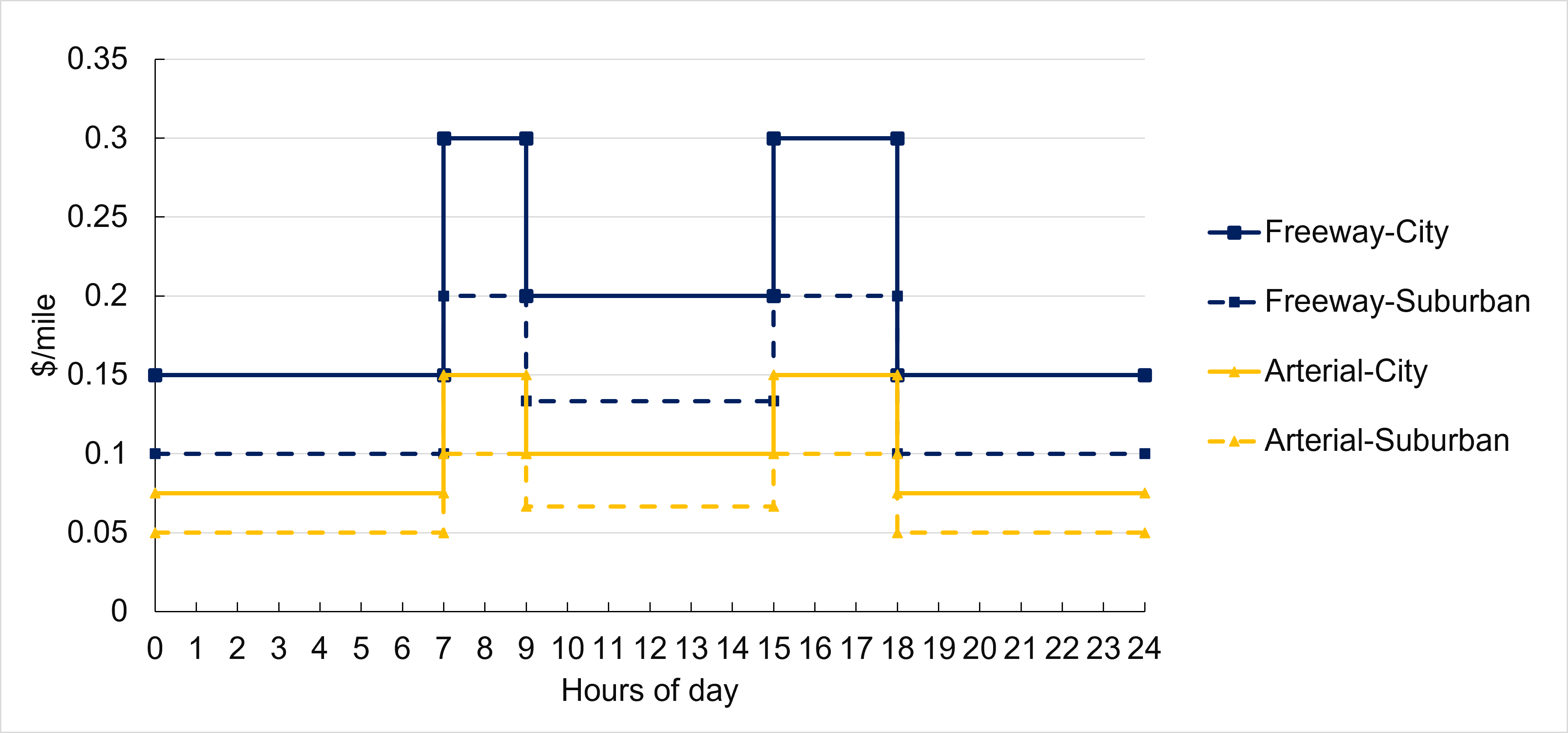}
    \caption{Smart pricing by location, functional, class, and time-of-day}
    \label{fig:vmt_tax}
\end{figure}

\begin{enumerate}
    \item Baseline (Scenario 01 BAU). The BAU scenario projects current trends to 2050 without any policy intervention, only incorporating 15.2\% telecommuting and low vehicle electrification (20.5\% of light-duty (LD) and 6.4\% of MDHD vehicles). This scenario serves as the primary baseline for all policy comparisons. Substantial transit improvements are planned for the year 2050, and they are incorporated into the BAU scenario \cite{CMAP2018ONTO2050}. These include the addition of several Bus Rapid Transit (BRT) lines for the Chicago Transit Authority (CTA) and stop-skipping bus routes (arterial rapid transit: ART) for the Pace Suburban Bus, extension of the CTA Red Line by four additional stations, extension of the METRA Commuter Rail's Union Pacific Northwest line by one station, and frequency increases across multiple agencies and routes resulting in a 37\% increase in overall revenue trips.
    \item Electrification (Scenarios 02a--02b). Two electrification scenarios examine the sensitivity of regional energy consumption to the pace of fleet electrification. Scenario 02a medium electrification reflects a moderate adoption trajectory, with 55.1\% of LD and 10.1\% of MDHD vehicles electrified. Scenario 02b high electrification represents an aggressive adoption pathway, reaching 94.0\% LD and 50.1\% MDHD electrification, consistent with projected EV penetration targets defined by the CMAP. All scenarios include freight dock charging (electric power connections at loading bays so freight vehicles can charge while parked, loading, or unloading.). The charging infrastructure for LD \cite{kvianipour2024trr} and MDHD \cite{kaleemExtremescaleEVCharging2026} vehicles were optimized for each scenario.
    \item Freight (Scenario 03 Freight Bundle). The freight bundle scenario evaluates a bundled freight demand management strategy. Our assumptions include combining off-hour delivery, a dedicated left lane for trucks, shore-powering idling trucks at the docks, and long-haul rail mode enforcement (assumed 50\% probability) for trips exceeding 500 miles. By deploying these four complementary interventions simultaneously, the scenario assesses their compounded effects relative to the BAU. The individual lever runs and their sensitivity analyses are omitted in this paper.
    \item Pricing (Scenarios 04a--04c). Three pricing scenarios examine distinct road-pricing strategies. Scenario 04a pricing expressway applies a facility-based toll targeting expressway corridors oriented toward the Chicago Loop. Scenario 04b pricing MFT implements a flat per-mile fee applied broadly across the network as a replacement for declining motor fuel tax (MFT) revenue as a result of electrification and vehicle efficiency improvements. Scenario 04c pricing smart deploys a spatiotemporally variable per-mile fee: freeways carry higher per-mile charges than arterials, urban city roads are priced above suburban and rural roads, and rates are further adjusted by time of day to discourage travel during peak congestion periods as shown in \Cref{fig:vmt_tax}. The low-income drivers are always charged the lowest rate (\$0.05 per mile) regardless of location or time-of-day.
    \item Transit (Scenario 05). The transit scenario represents a high-investment future on top of the already-improved BAU \cite{CMAP2018ONTO2050}. On the rail side, CTA extensions bring the Green Line to Midway and Jackson Park, the Blue Line to Mannheim, the Brown Line to Jefferson Park, and the Yellow Line to Old Orchard. Metra commuter rail expansion includes the Milwaukee North and Milwaukee West extensions. On the bus side, a CTA Halsted BRT corridor is added, and a 20\% speed improvement is applied across 48 routes. The frequencies are multiplied by two, resulting in 101\% increase in overall revenue trips with the compound effect of route additions.
\end{enumerate}

\section{Results}\label{sec:results}
\begin{figure}[!htbp]
    \centering
    \includegraphics[width=\linewidth]{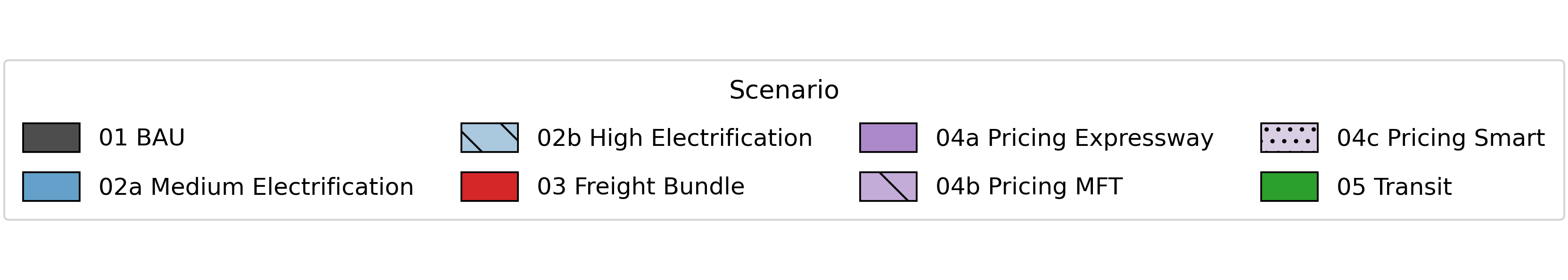}
    \caption{Scenario plot legend.}
    \label{fig:scenario_legend}
\end{figure}

This section presents the impacts of the modeled policy and technology scenarios on travel behavior, system performance, user costs, and energy demand in the Chicago metropolitan region by 2050. All results are reported relative to the BAU baseline. All figures use a consistent color scheme for each scenario, as shown in \Cref{fig:scenario_legend}. This legend is omitted from individual plots to avoid repetition.

\subsection{System-Level Trip Metrics}
\begin{figure}[!htbp]
    \centering

    \begin{subfigure}{0.32\textwidth}
        \centering
        \includegraphics[width=\linewidth]{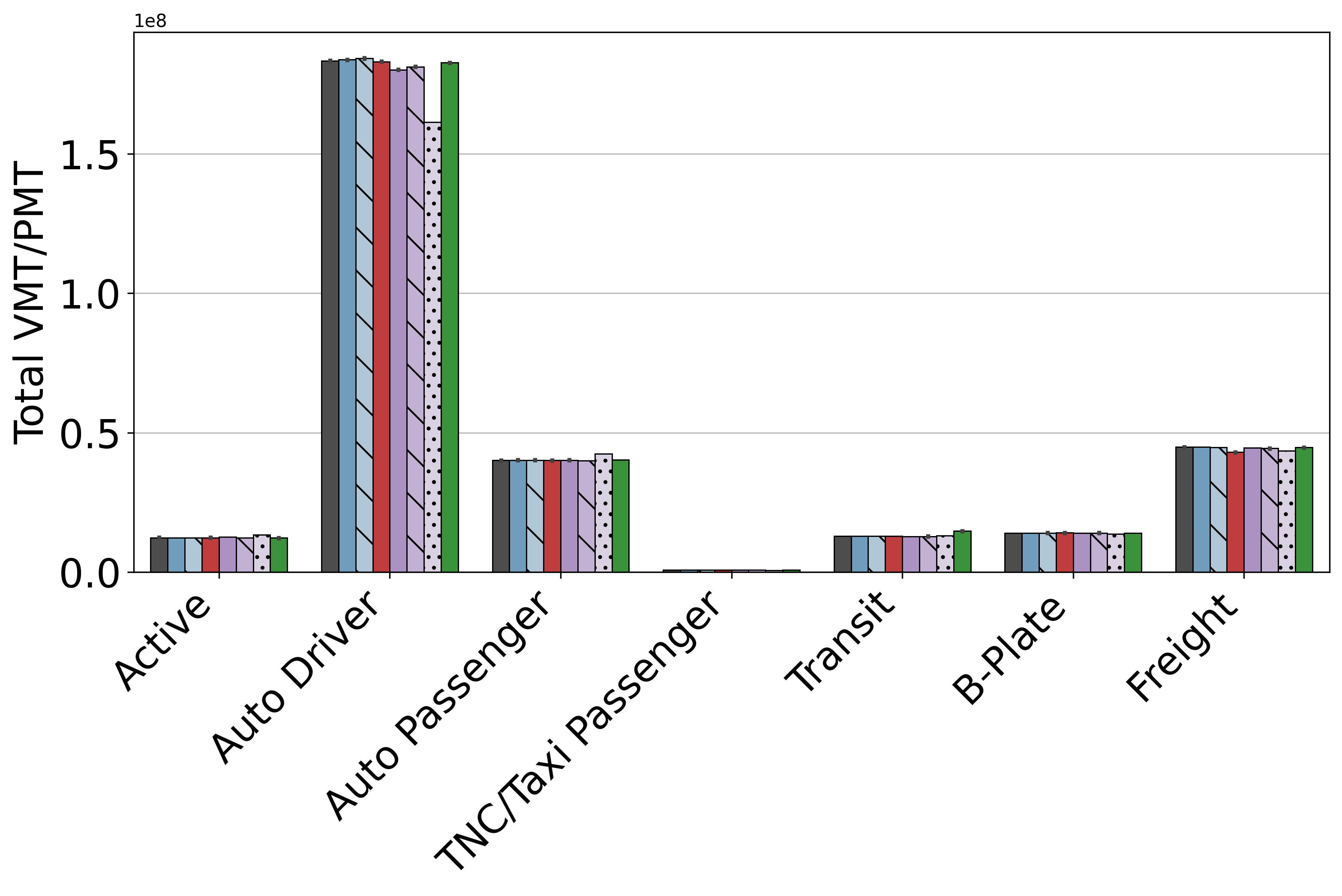}
        \caption{Total VMT/PMT by mode.}
        \label{fig:vmt_vht_total_vmt}
    \end{subfigure}
    \hspace{0.03\textwidth}
    \begin{subfigure}{0.32\textwidth}
        \centering
        \includegraphics[width=\linewidth]{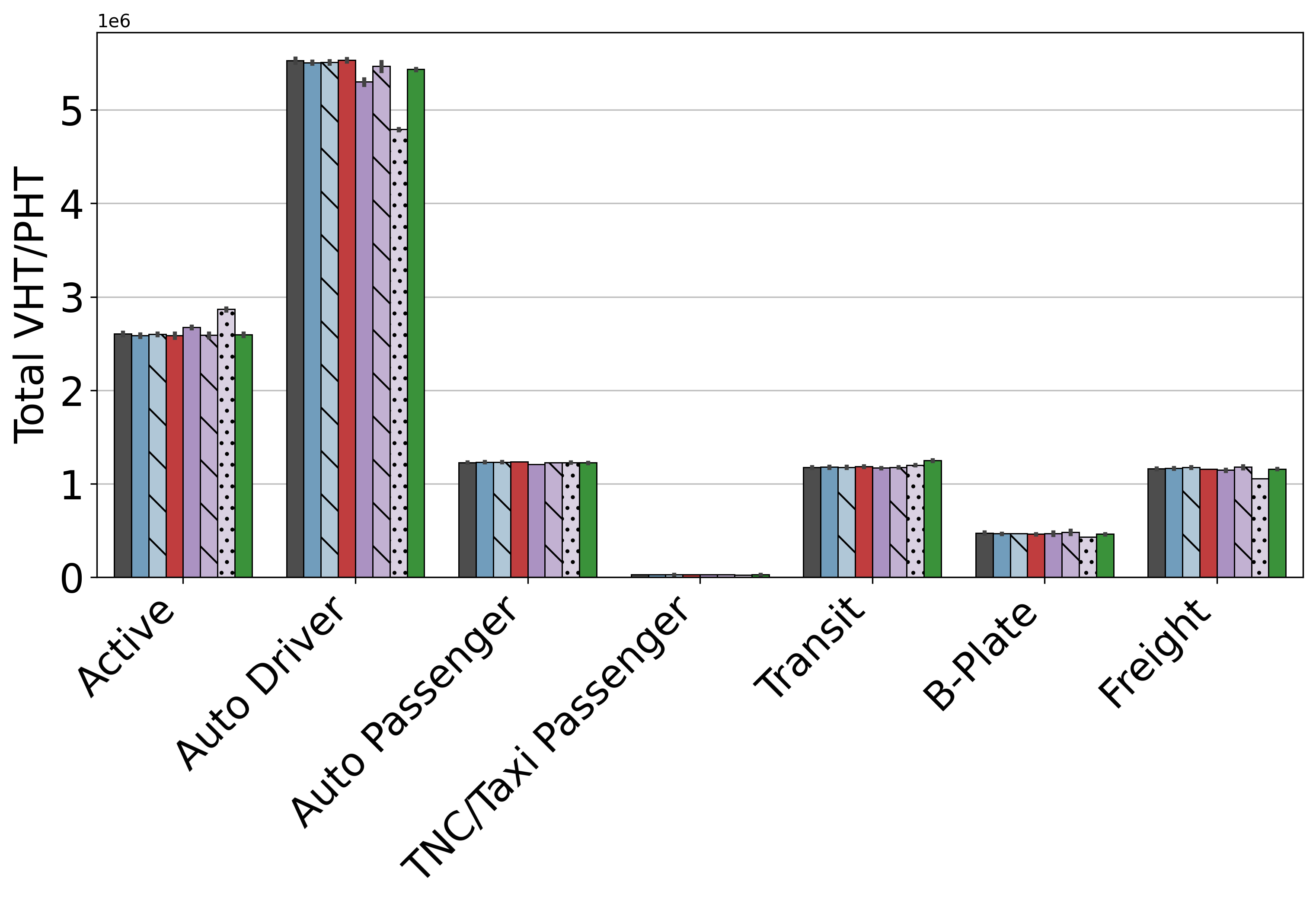}
        \caption{Total VHT/PHT by mode.}
        \label{fig:vmt_vht_total_vht}
    \end{subfigure}

    \vspace{0.5em}

    \begin{subfigure}{0.32\textwidth}
        \centering
        \includegraphics[width=\linewidth]{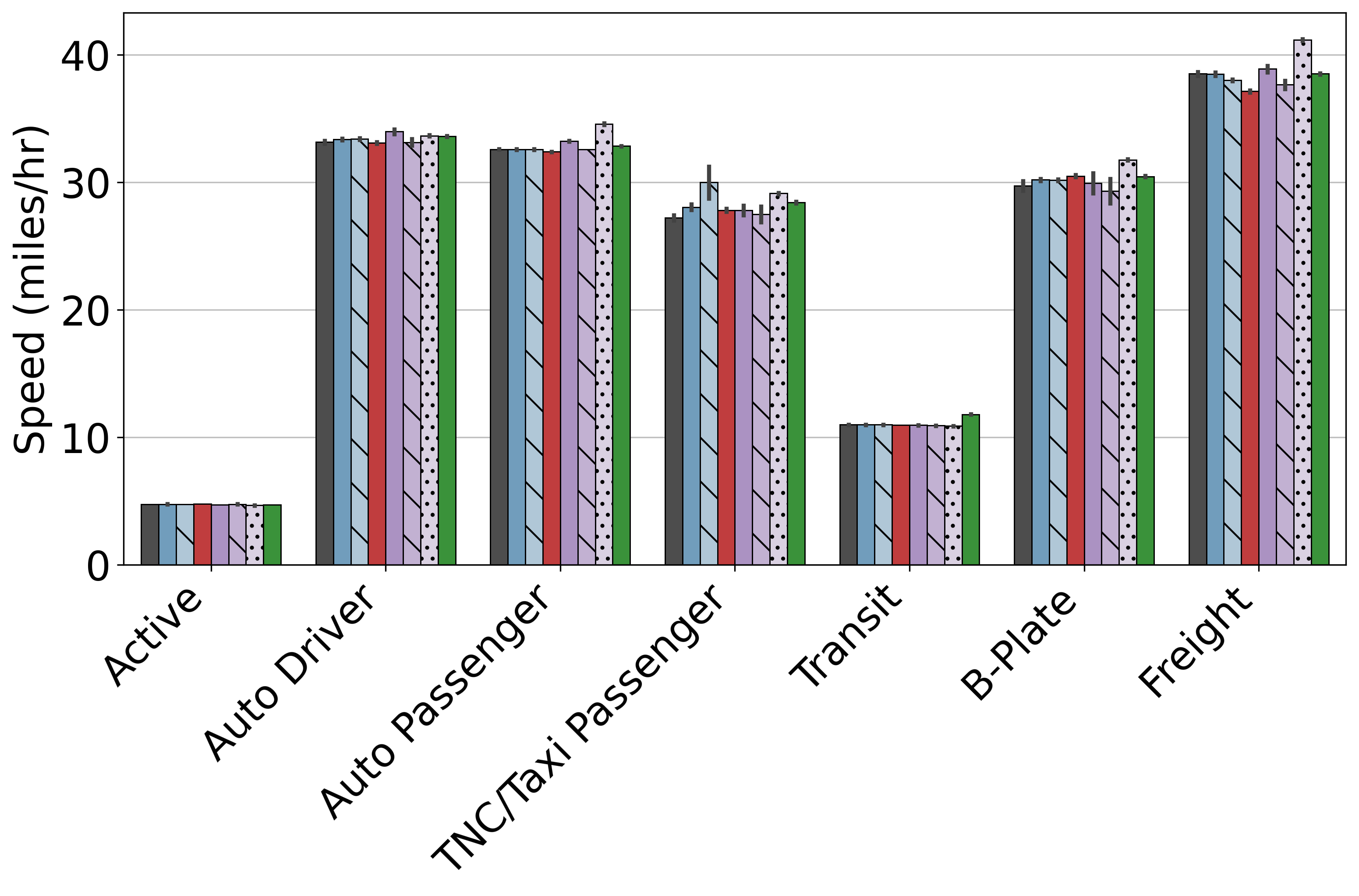}
        \caption{Average speed by mode.}
    \end{subfigure}
    \hfill
    \begin{subfigure}{0.32\textwidth}
        \centering
        \includegraphics[width=\linewidth]{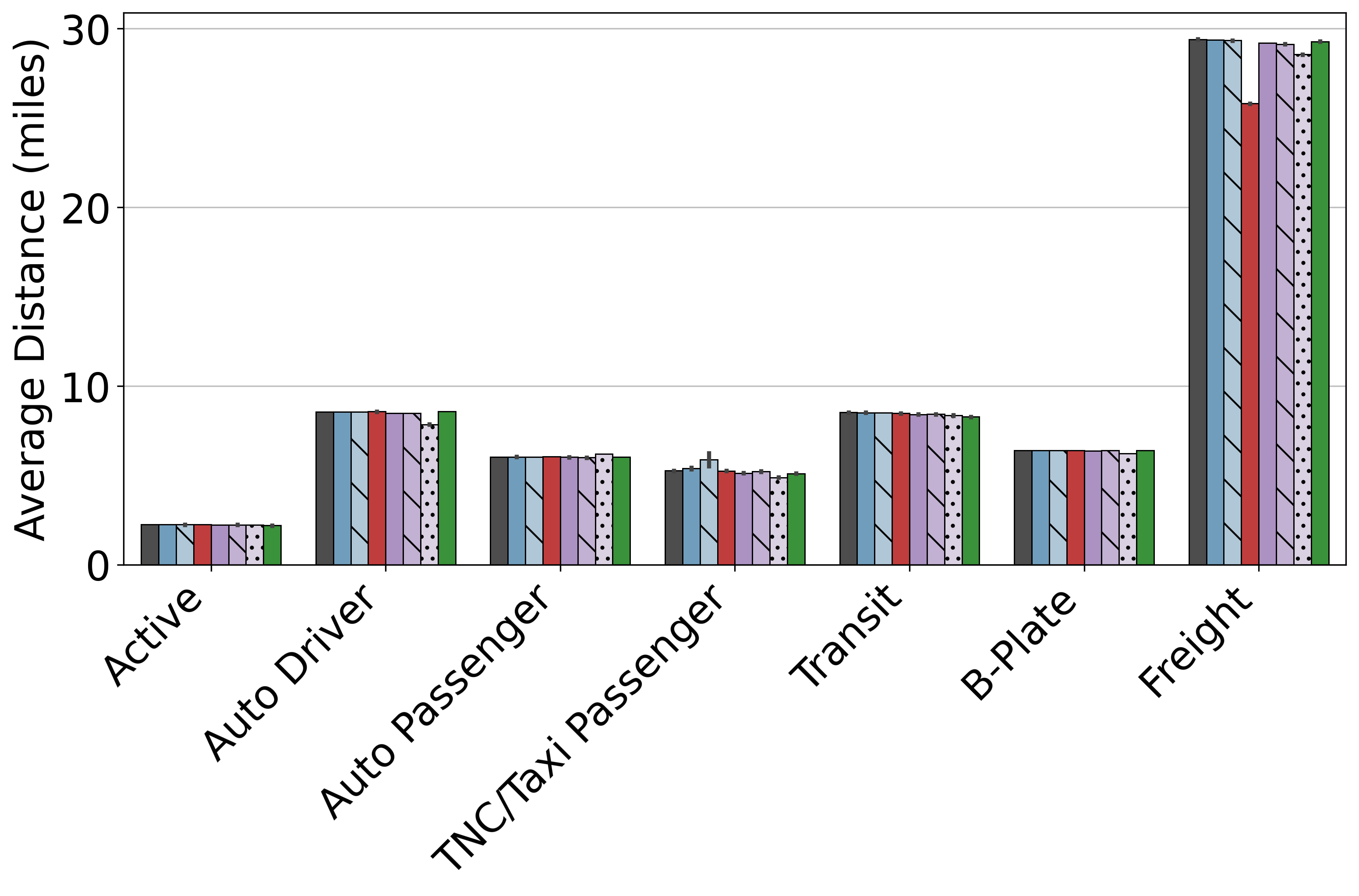}
        \caption{Average distance by mode.}
    \end{subfigure}
    \hfill
    \begin{subfigure}{0.32\textwidth}
        \centering
        \includegraphics[width=\linewidth]{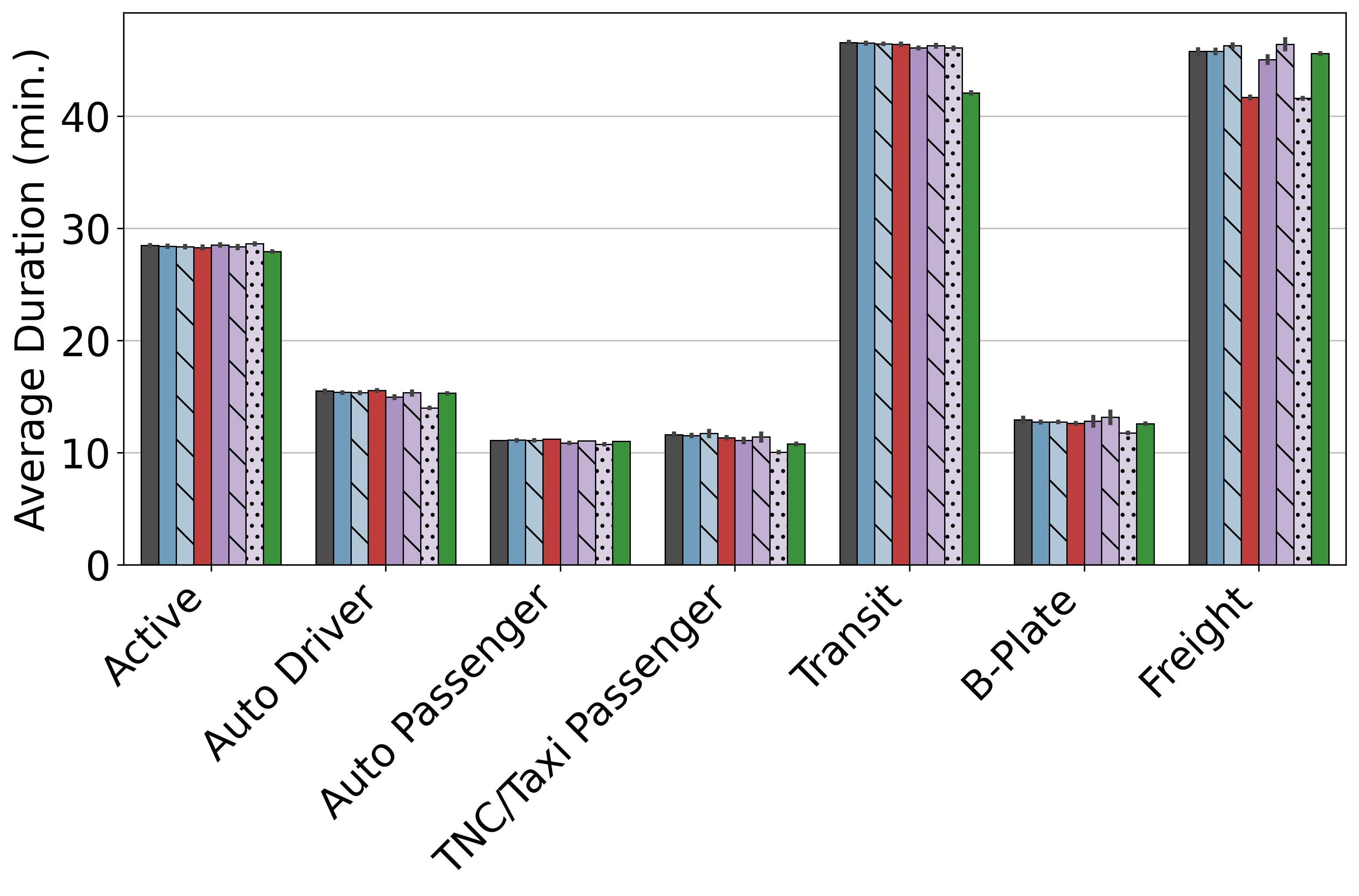}
        \caption{Average duration by mode.}
        \label{fig:vmt_vht_avg_duration}
    \end{subfigure}

    \caption{Trip metrics by modes across all scenarios relative to the BAU baseline.}
    \label{fig:resnet_metrics}
\end{figure}

\Cref{fig:resnet_metrics} compares total and average trip metrics—including VMT, VHT, PMT, PHT, average speed, and trip duration—across all scenarios and travel modes. Active, auto driver, auto passenger, TNC/taxi passenger, and transit values are reported in PMT and PHT. On the other hand freight and service (B-plate) travel is reported in VMT and VHT. Electrification scenarios produce negligible changes in VMT and VHT, as expected, since fleet electrification does not alter travel demand. However, at the trip level, the high electrification scenario is associated with a notable increase in TNC average speed and trip distance, suggesting longer and faster TNC trips under that configuration.

Most policy interventions produce only slight deviations from the BAU baseline in aggregate VMT/PMT and VHT/PHT, reflecting the inherent inertia of automobile dependence even under significant policy reform. However, the smart pricing scenario stands out as the most effective lever for reducing auto driver PMT and PHT, with a 10\% reduction in trip duration compared to BAU (\Cref{fig:vmt_vht_total_vmt,fig:vmt_vht_total_vht,fig:vmt_vht_avg_duration}). This is consistent with a behavioral response in which higher costs discourage longer discretionary auto trips.

\subsection{Freight Impacts}
\begin{figure}[!htbp]
    \centering

    \begin{subfigure}[t]{0.31\textwidth}
        \centering
        \includegraphics[width=\linewidth]
        {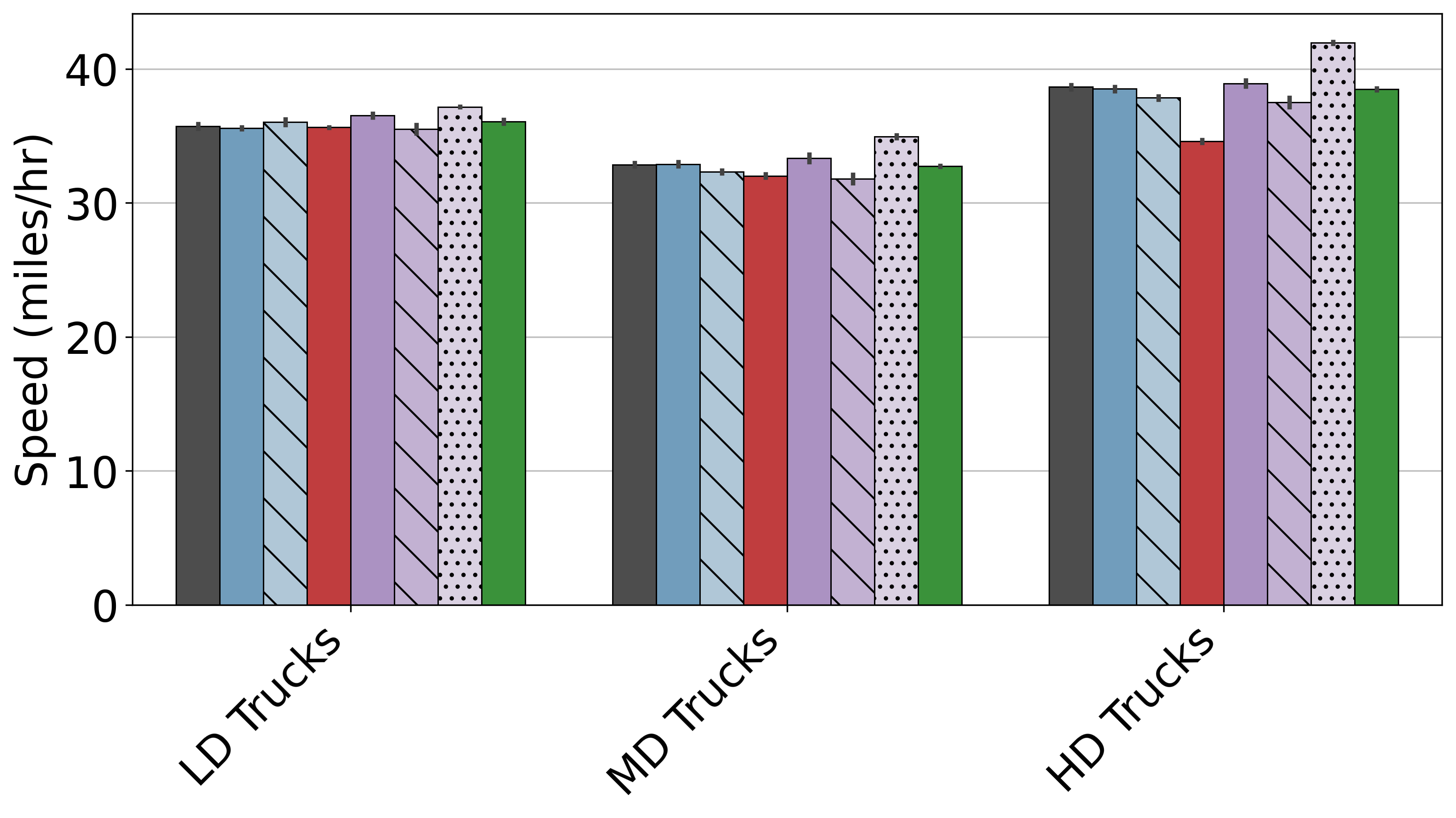}
        \caption{Average speed by freight mode.}
        \label{fig:freight_average_speed}
    \end{subfigure}
    \hfill
    \begin{subfigure}[t]{0.31\textwidth}
        \centering
        \includegraphics[width=\linewidth]
        {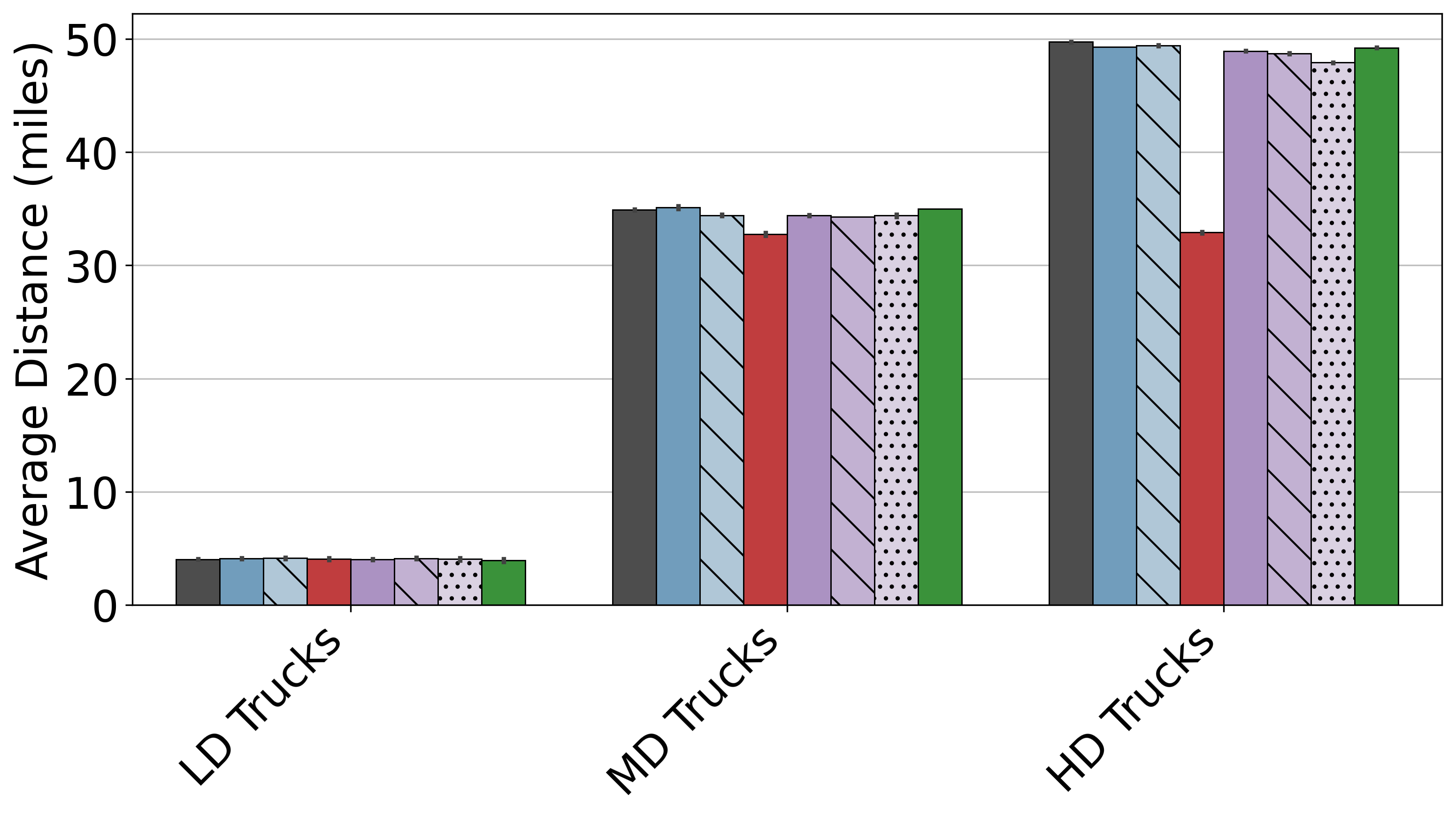}
        \caption{Average distance by freight mode.}
        \label{fig:freight_average_distance}
    \end{subfigure}
    \hfill
    \begin{subfigure}[t]{0.31\textwidth}
        \centering
        \includegraphics[width=\linewidth]
        {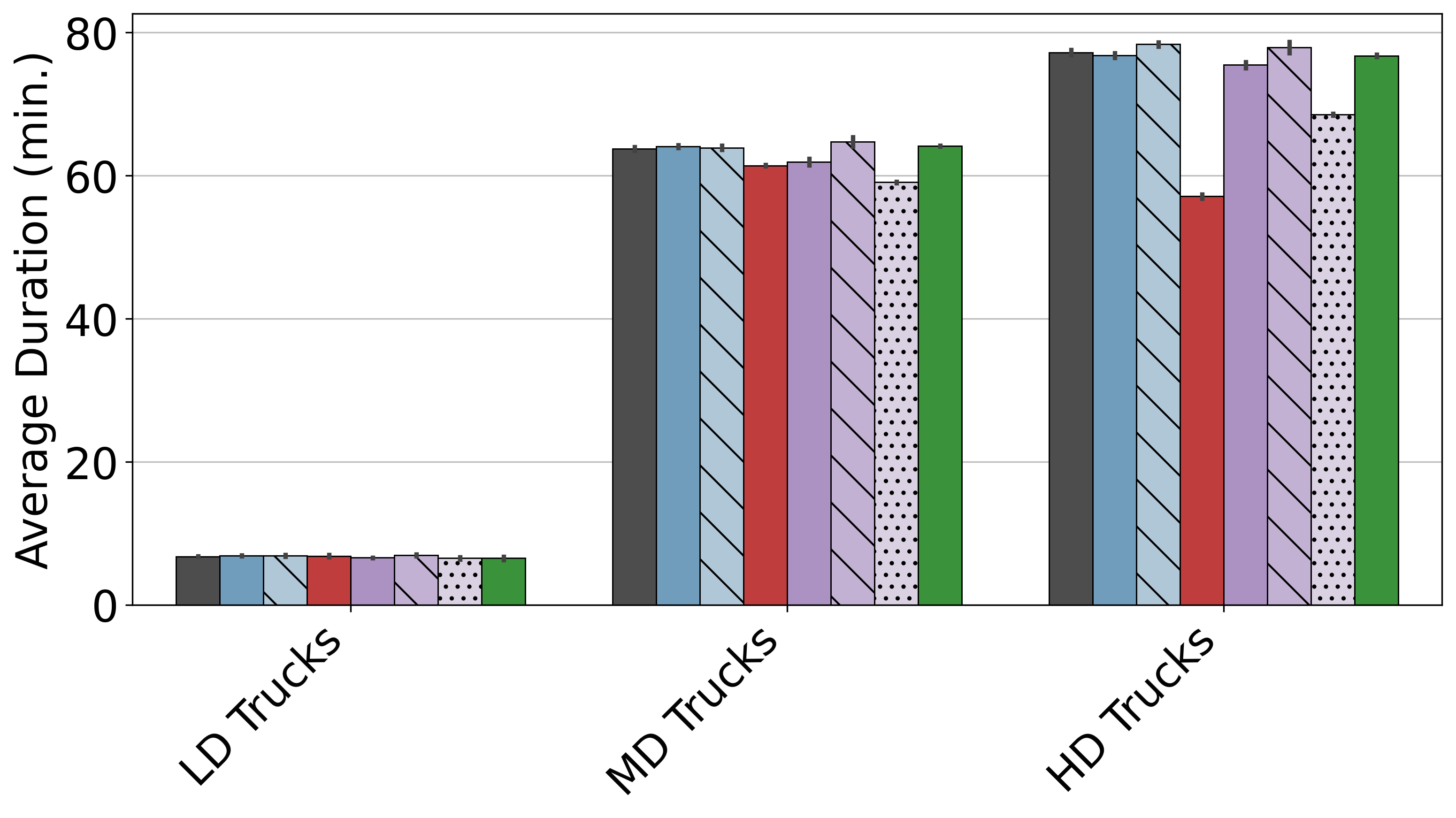}
        \caption{Average duration by freight mode.}
        \label{fig:freight_average_duration}
    \end{subfigure}

    \vspace{0.5em}

    \begin{subfigure}[t]{0.31\textwidth}
        \centering
        \includegraphics[width=\linewidth]
        {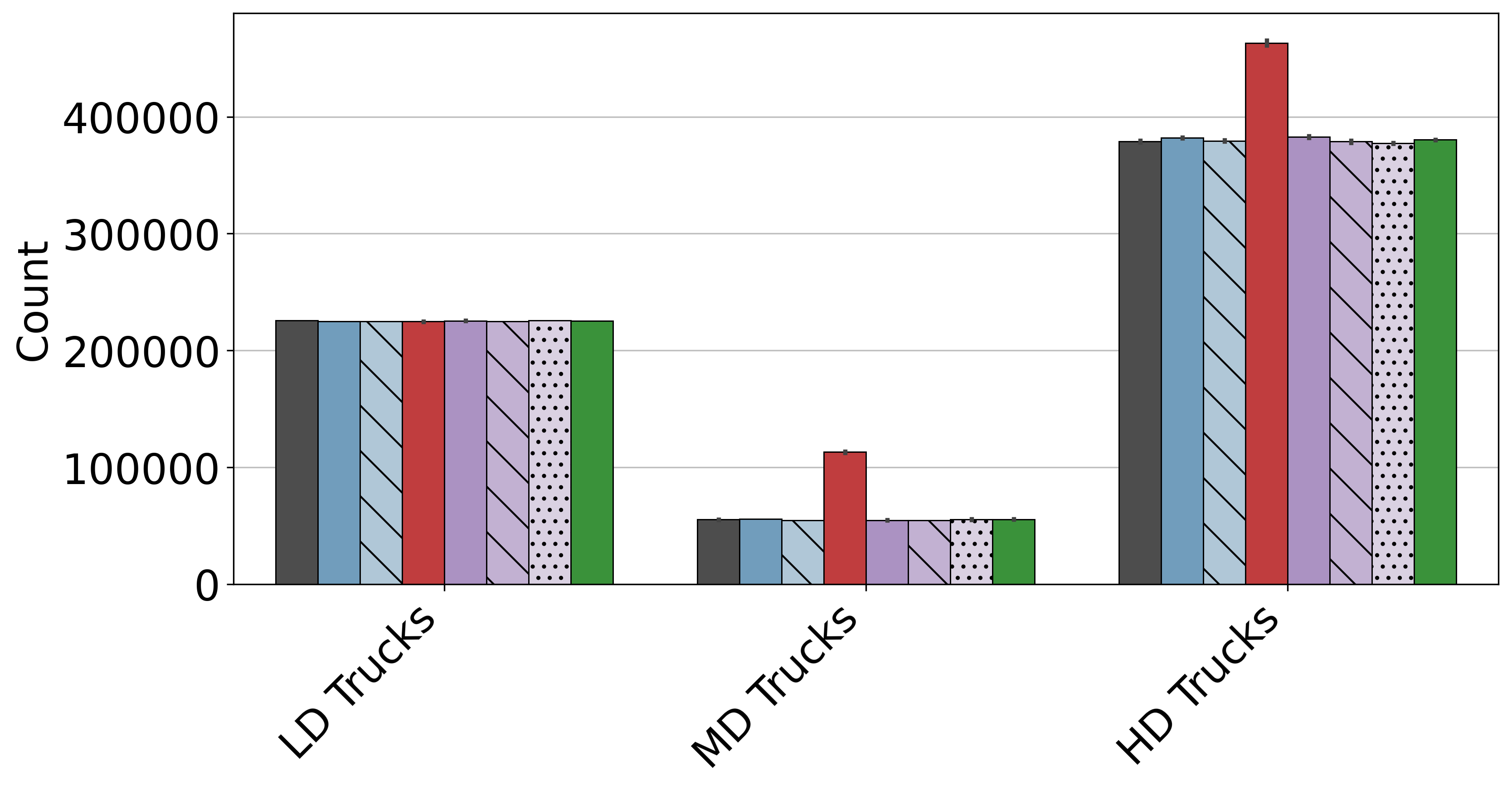}
        \caption{Trip count by freight mode.}
        \label{fig:freight_trip_count}
    \end{subfigure}
    \hfill
    \begin{subfigure}[t]{0.31\textwidth}
        \centering
        \includegraphics[width=\linewidth]
        {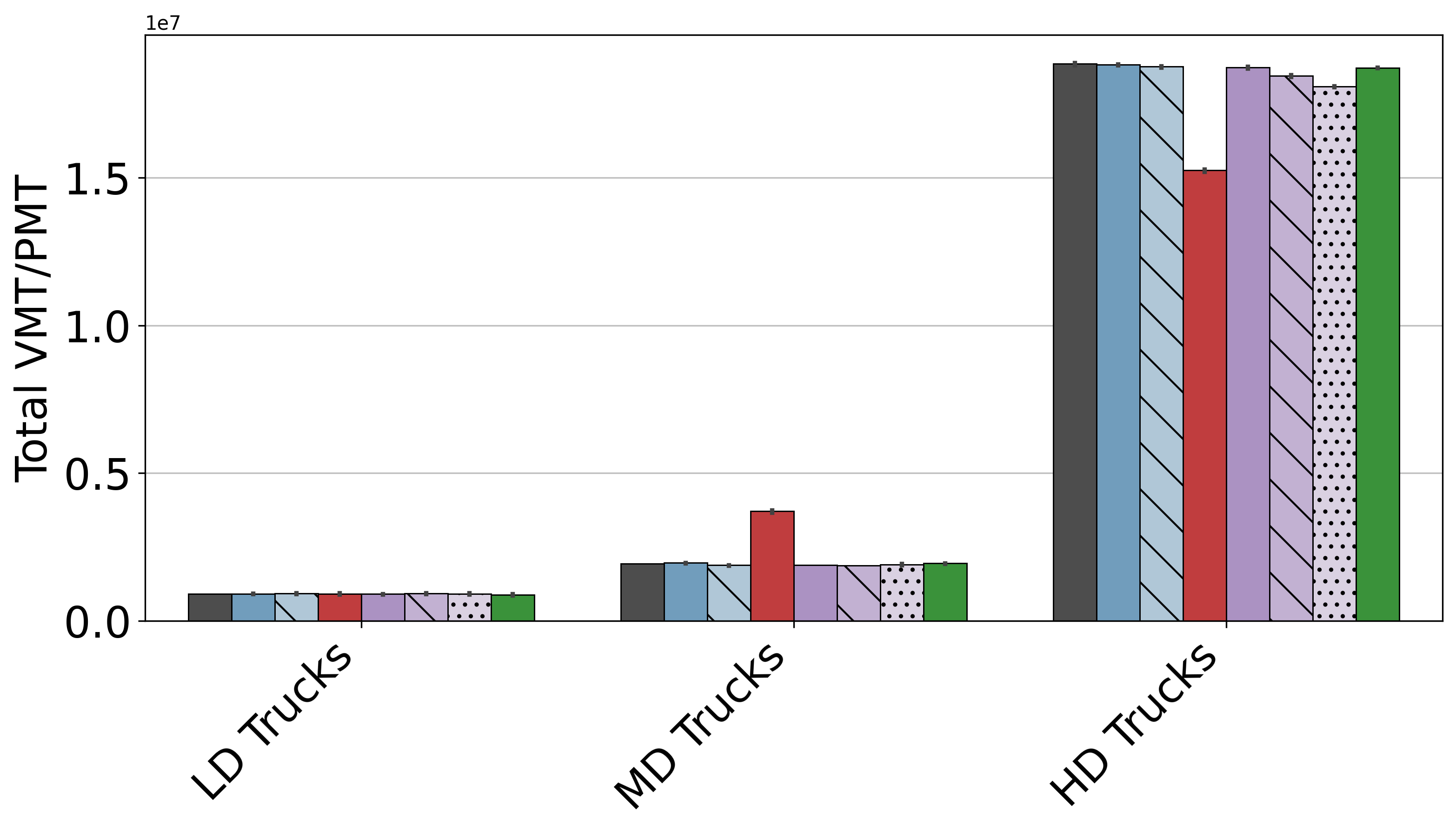}
        \caption{Total VMT by freight mode.}
        \label{fig:freight_total_vmt}
    \end{subfigure}
    \hfill
    \begin{subfigure}[t]{0.31\textwidth}
        \centering
        \includegraphics[width=\linewidth]
        {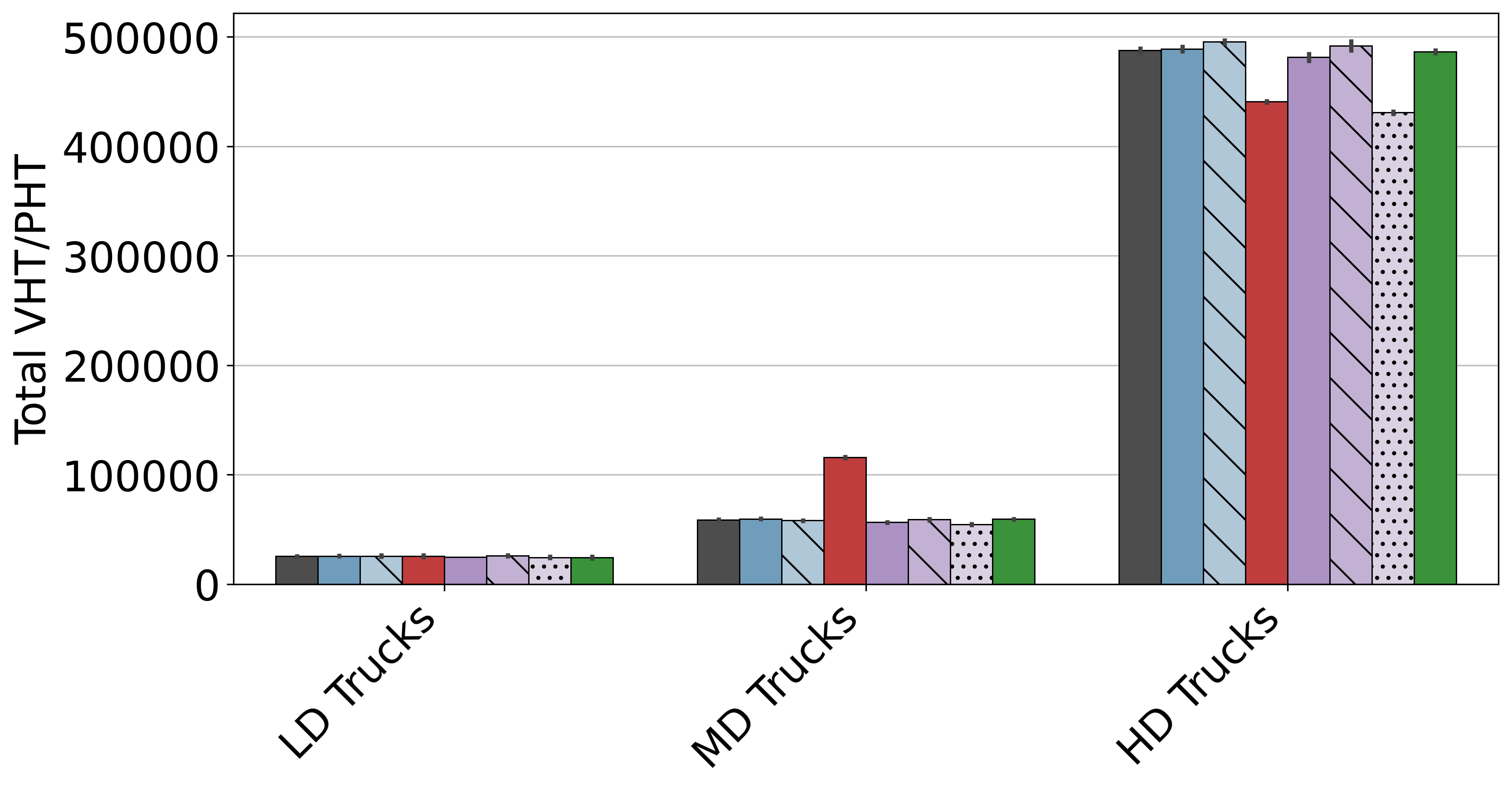}
        \caption{Total VHT by freight mode.}
        \label{fig:freight_total_vht}
    \end{subfigure}

    \caption{Trip metrics by freight modes across all scenarios relative to the BAU baseline.}
    \label{fig:trip_metrics_freight_modes}
\end{figure}

Freight VMT remains broadly stable across most scenarios, with the notable exception of the Freight Bundle scenario, which produces a measurable reduction in VMT accompanied by a decline in average freight speed of approximately 8\% as shown in \Cref{fig:freight_average_speed}. This likely reflects the operational constraints of off-hour delivery windows rather than a reduction in freight demand per se.

The Freight Bundle scenario increases the number of truck trips while simultaneously reducing total VMT across all truck classes. This apparent paradox is resolved by the trip-level metrics: the Freight Bundle produces shorter and slower individual trips, meaning more trips are made but each covers less distance, resulting in a net reduction in system-level freight VMT. Looking at truck counts, the Freight Bundle stands out most prominently in the medium-duty truck category, where trip counts rise substantially above the BAU baseline, while heavy-duty truck counts also increase modestly. This increase in trip frequency is consistent with the off-hour delivery component of the bundle, which may fragment consolidated shipments into more frequent and smaller deliveries. Additionally, the rail diversion policy, which reassigns long-haul movements to rail, creates shorter drayage trips for trucks. Therefore, there is more VMT and VHT for medium-duty trucks, and less for heavy-duty trucks. The Freight Bundle produces the lowest average trip speed across all three truck classes, approximately 31 mph for MDHD trucks compared to the BAU values of around 33–39 mph, consistent with the operational friction introduced by dedicated truck lanes and off-hour delivery constraints. Average trip distance under the Freight Bundle is shorter (28 miles compared to BAU), likely due to the rail diversion of trips exceeding 500 miles.

\subsection{Pricing Impacts}
\begin{figure}[!ht]
    \centering

    \begin{subfigure}[t]{0.48\textwidth}
        \centering
        \includegraphics[width=\linewidth]{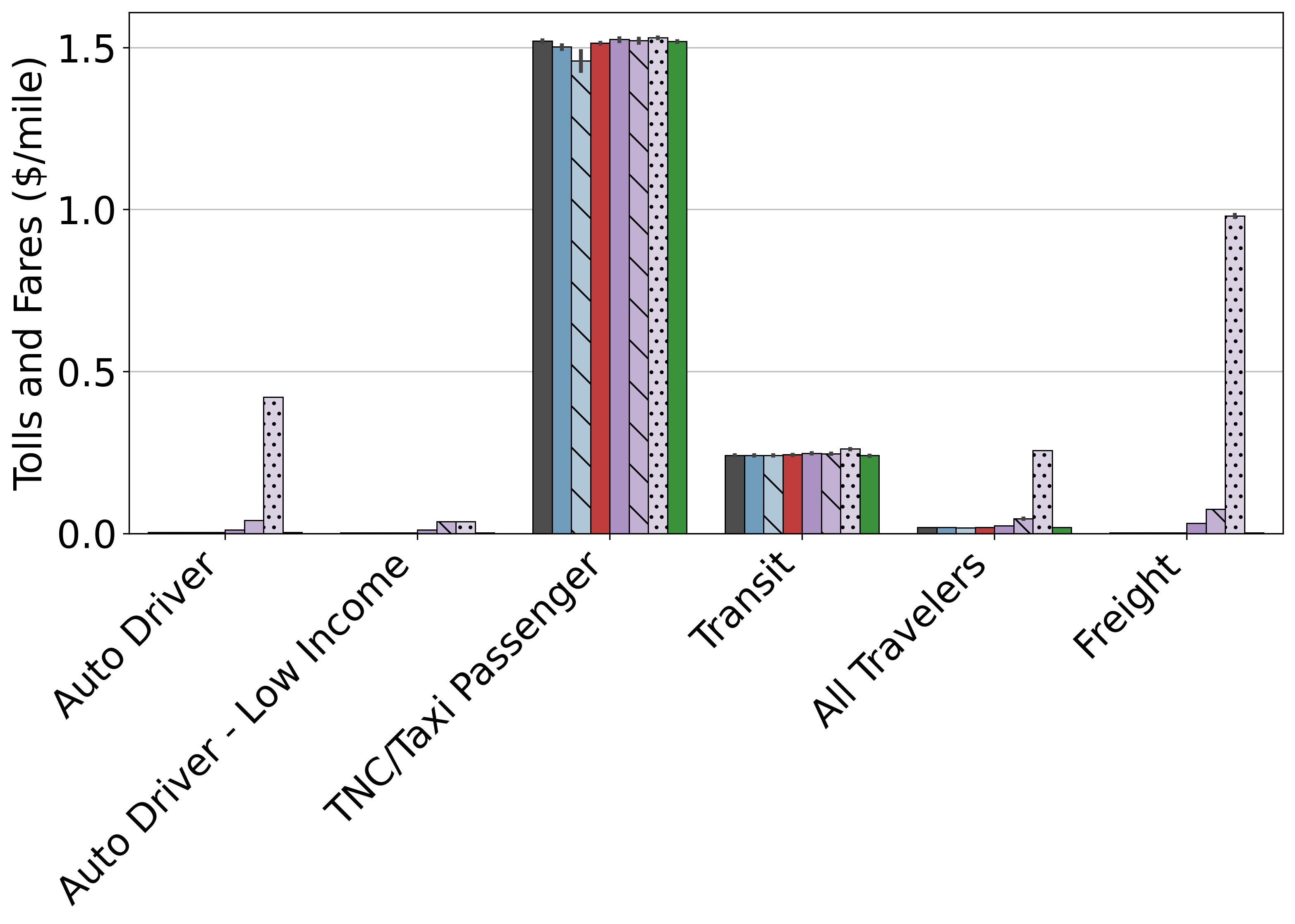}
        \caption{Tolls and fares (\$/mile).}
        \label{fig:trip_costs_cost_per_mile_tolls_fares}
    \end{subfigure}
    \hfill
    \begin{subfigure}[t]{0.48\textwidth}
        \centering
        \includegraphics[width=\linewidth]{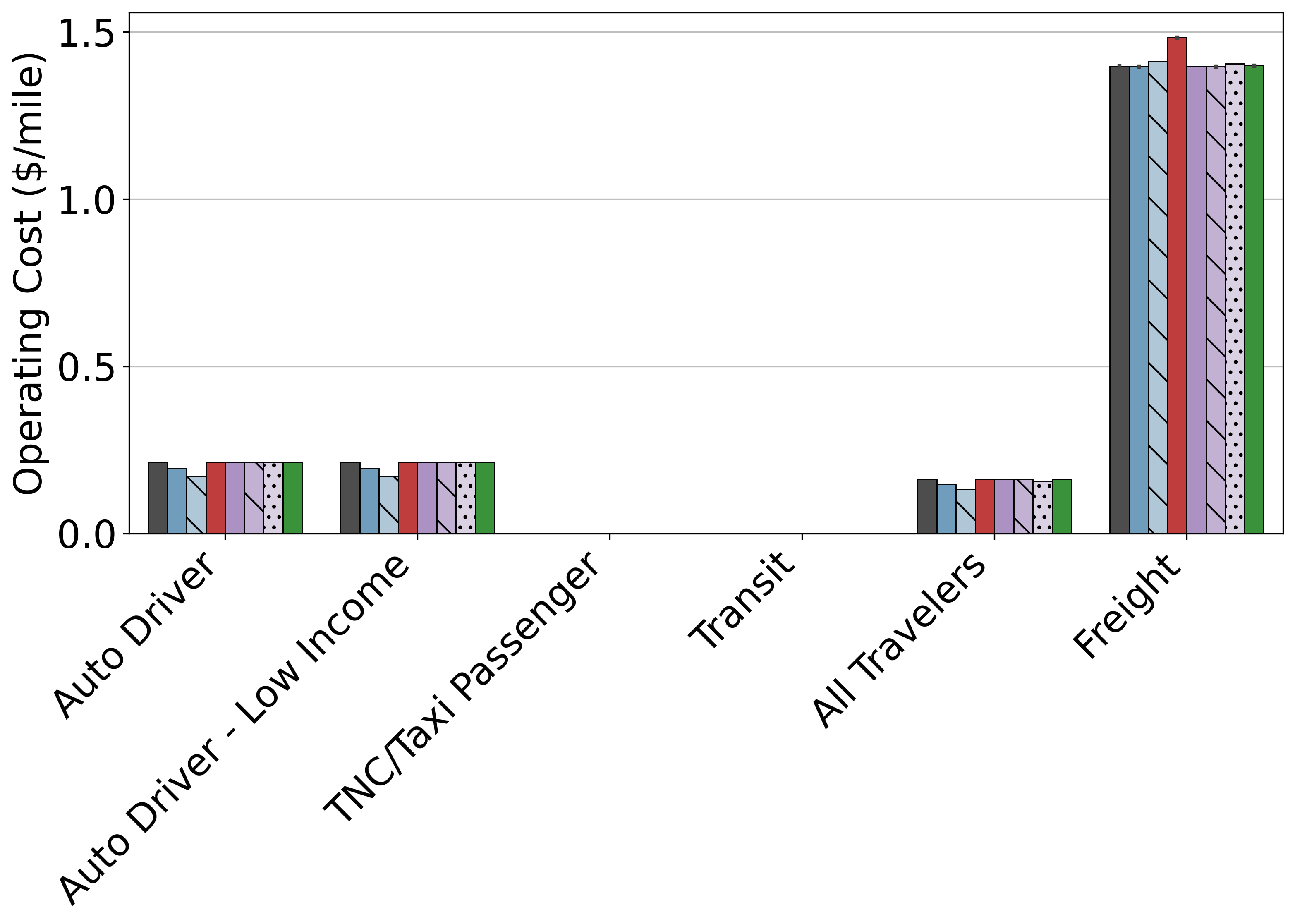}
        \caption{Operating cost (\$/mile).}
    \end{subfigure}

    \vspace{0.5em}

    \begin{subfigure}[t]{0.48\textwidth}
        \centering
        \includegraphics[width=\linewidth]{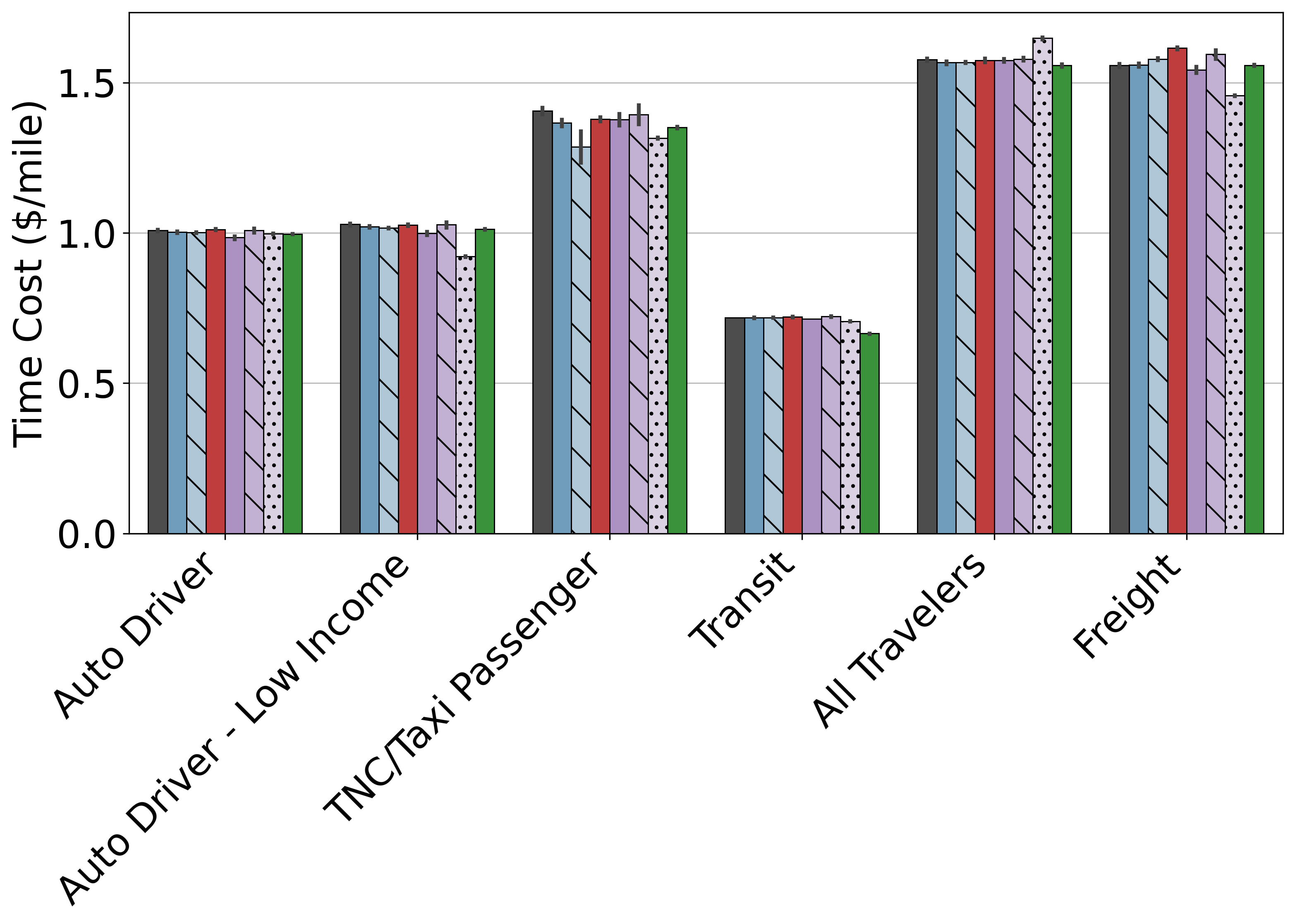}
        \caption{Time cost (\$/mile).}
    \end{subfigure}
    \hfill
    \begin{subfigure}[t]{0.48\textwidth}
        \centering
        \includegraphics[width=\linewidth]{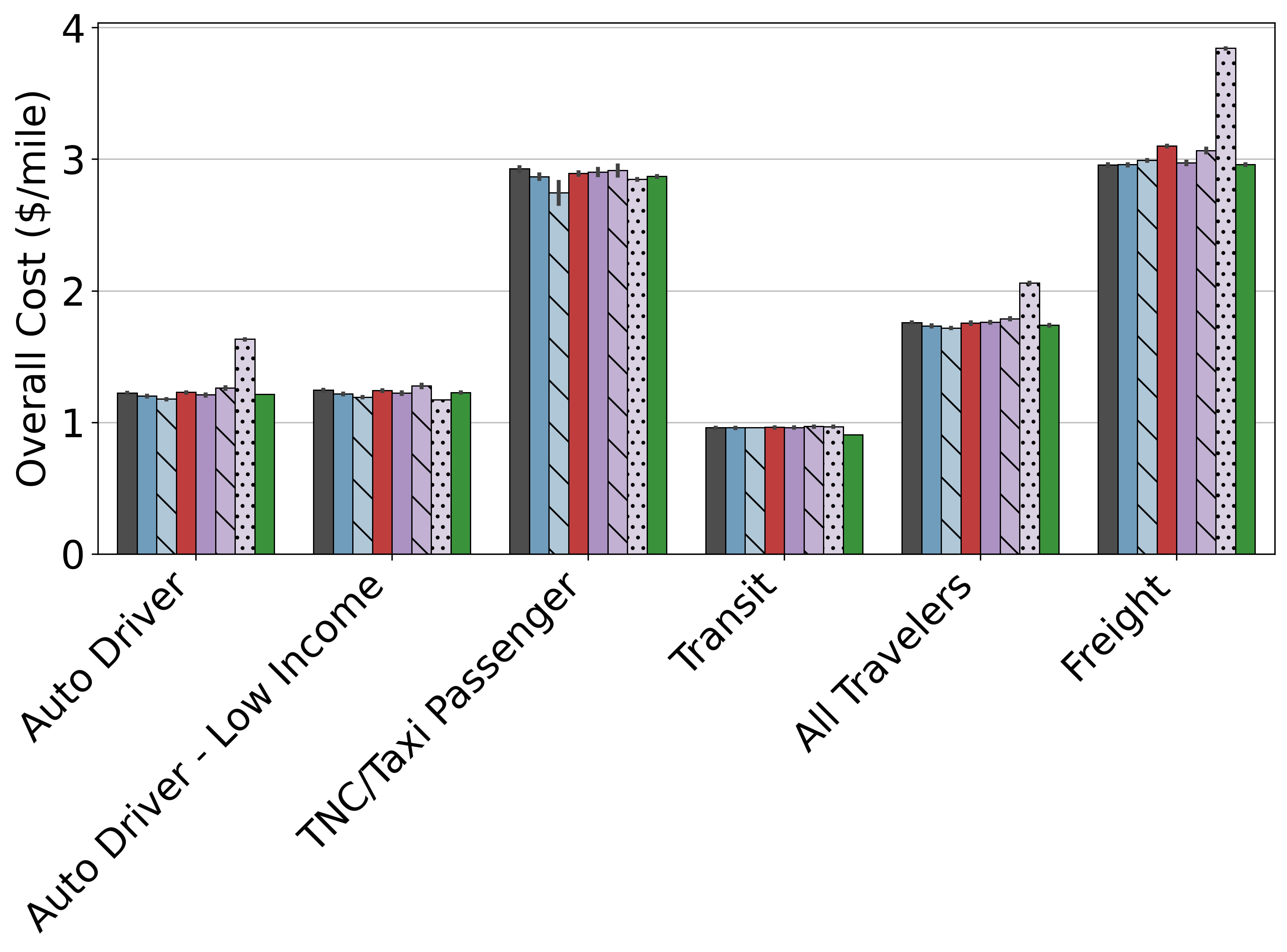}
        \caption{Overall cost (\$/mile).}
    \end{subfigure}

    \caption{Cost breakdown by modes across all scenarios relative to the BAU baseline.}
    \label{fig:cost_metrics_modes}
\end{figure}

The cost metrics further illuminate the mechanisms driving behavioral change. As shown in \Cref{fig:trip_costs_cost_per_mile_tolls_fares}, under smart pricing, tolls and fares for auto drivers rise substantially (\$0.5/mile) compared to other scenarios (<\$0.05/mile). TNC/taxi passengers bear the highest per-mile toll burden (approximately \$1.55/mile), nearly constant across all scenarios. Operating costs and time cost remain relatively stable across scenarios within each mode. This indicates that the primary cost signal for driving behavioral change of a specific mode is the pricing instrument itself rather than vehicle operating expenses and time costs. Time costs across all traveler categories are broadly similar across scenarios (\$1.00--\$1.10/mile for auto drivers), though freight exhibits notably higher time costs (\$1.60--\$1.75/mile). In terms of overall cost per mile, smart pricing results in among the highest total burdens for auto drivers and TNC/Taxi passengers, reinforcing why this scenario produces the strongest suppression of automobile travel demand. 

\subsection{Transit System Performance}
The transit scenario produces a substantial increase in transit VMT of approximately 14\%, paired with increases in average transit speed. The transit expansion and speed improvements attract longer-distance riders. \Cref{tab:transit_performance} summarizes four system-level performance indicators for the BAU, smart pricing and transit expansion scenarios.

Under the BAU, the system serves 2,175,455 daily riders. The transit expansion scenario increases ridership by nearly 18\% relative to BAU. This magnitude of ridership growth underscores the transformative demand potential of coordinated network investment, particularly the rail extensions, BRT corridors, and speed improvements embedded in this scenario.

Relative to the BAU, the transit expansion scenario increases system PMT by 14\% and PHT by 6\%, reflecting the additional vehicle-miles and vehicle-hours required to serve the expanded network and more trips. 

The average origin-to-destination travel speed under BAU is 11 mph and improves to 12 mph under the transit expansion scenario. This is a 7\% gain from 20\% speed enhancement that was applied to 48 existing routes, BRT corridor investments, and the addition of higher-speed rail extensions. All these enhancements shift the fleet composition toward faster, higher-capacity services.

These results demonstrate that ambitious transit investments can yield substantial increases in ridership and service coverage, and can improve overall system speed. This suggests that coordinated, high-capacity transit expansion is a robust strategy for meeting future mobility needs in the Chicago region.

Additionally, the smart pricing scenario produces a 2.5\% increase in transit ridership relative to BAU. Besides, we observe a 1.0\% increase in passenger-miles and 2.0\% increase in passenger-hours, while average speed remains unchanged at 11~mph. The auto cost disincentives under smart pricing may generate a slight mode shift toward transit.

\begin{table}[ht]
\centering
\caption{Transit System Performance Indicators by Scenario}
\label{tab:transit_performance}
\begin{threeparttable}
\scriptsize
\begin{tabular}{lrrrrr}
\toprule
\textbf{Performance Indicator}
    & \textbf{01 BAU}
    & \textbf{04c Pricing Smart}
    & \textbf{Change}
    & \textbf{05 Transit}
    & \textbf{Change} \\
\midrule
Total Ridership (daily)
    & 2,175,455
    & 2,228,766
    & $+$2.5\%
    & 2,576,684
    & $+$18.4\% \\
Total Passenger-miles
    & 12,944,755
    & 13,075,633
    & $+$1.0\%
    & 14,757,937
    & $+$14.0\% \\
Total Passenger-hours
    & 1,177,433
    & 1,200,674
    & $+$2.0\%
    & 1,249,290
    & $+$6.1\% \\
Avg.\ Origin-Destination Speed (mph)
    & 11
    & 11
    & $\pm$0\%
    & 12
    & $+$7\% \\
\bottomrule
\end{tabular}
\end{threeparttable}
\end{table}

\begin{figure}[!htbp]
    \centering

    \begin{subfigure}[t]{0.31\textwidth}
        \centering
        \includegraphics[width=\linewidth]
        {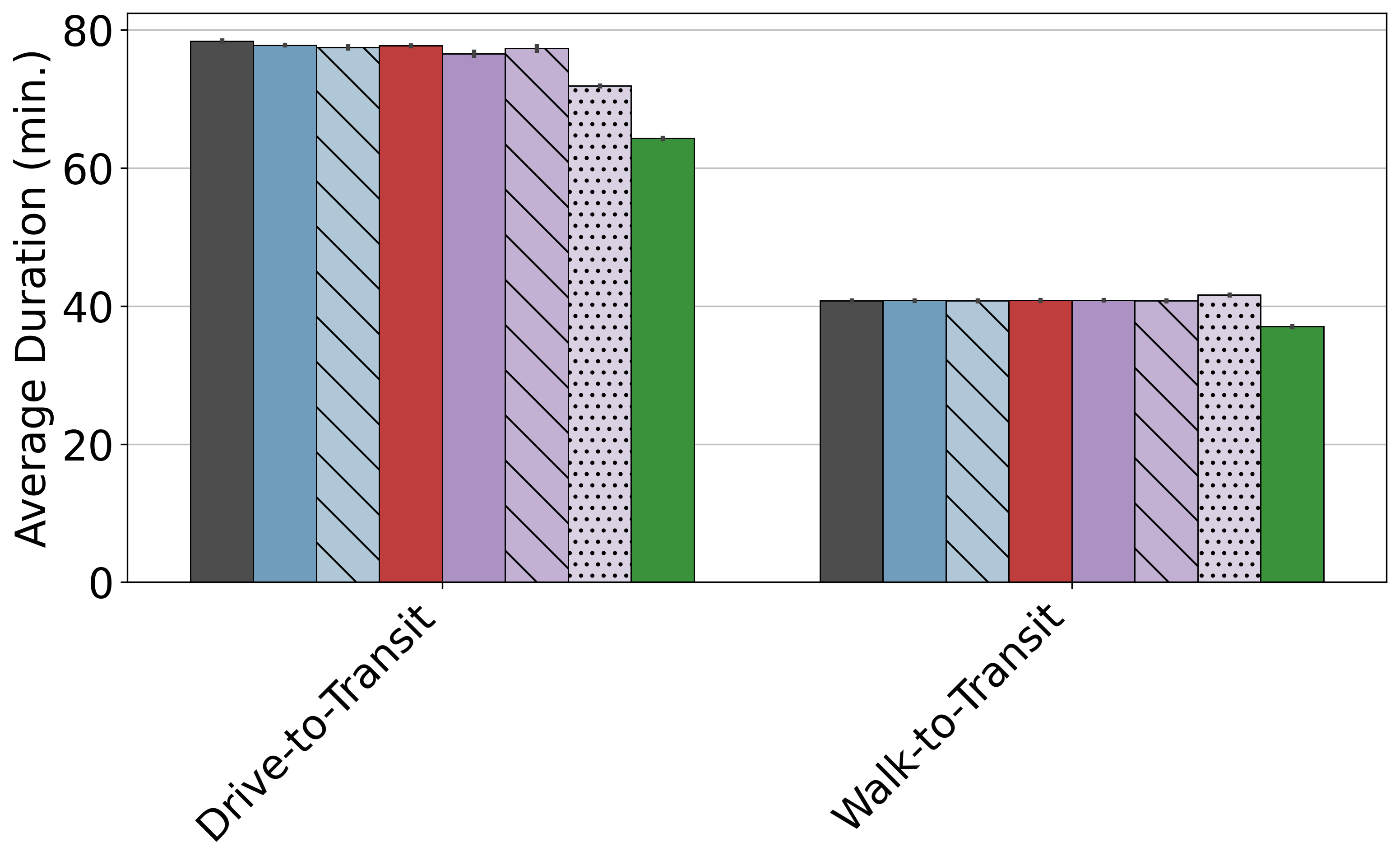}
        \caption{Average duration by access mode.}
        \label{fig:transit_avg_duration}
    \end{subfigure}
    \hfill
    \begin{subfigure}[t]{0.31\textwidth}
        \centering
        \includegraphics[width=\linewidth]
        {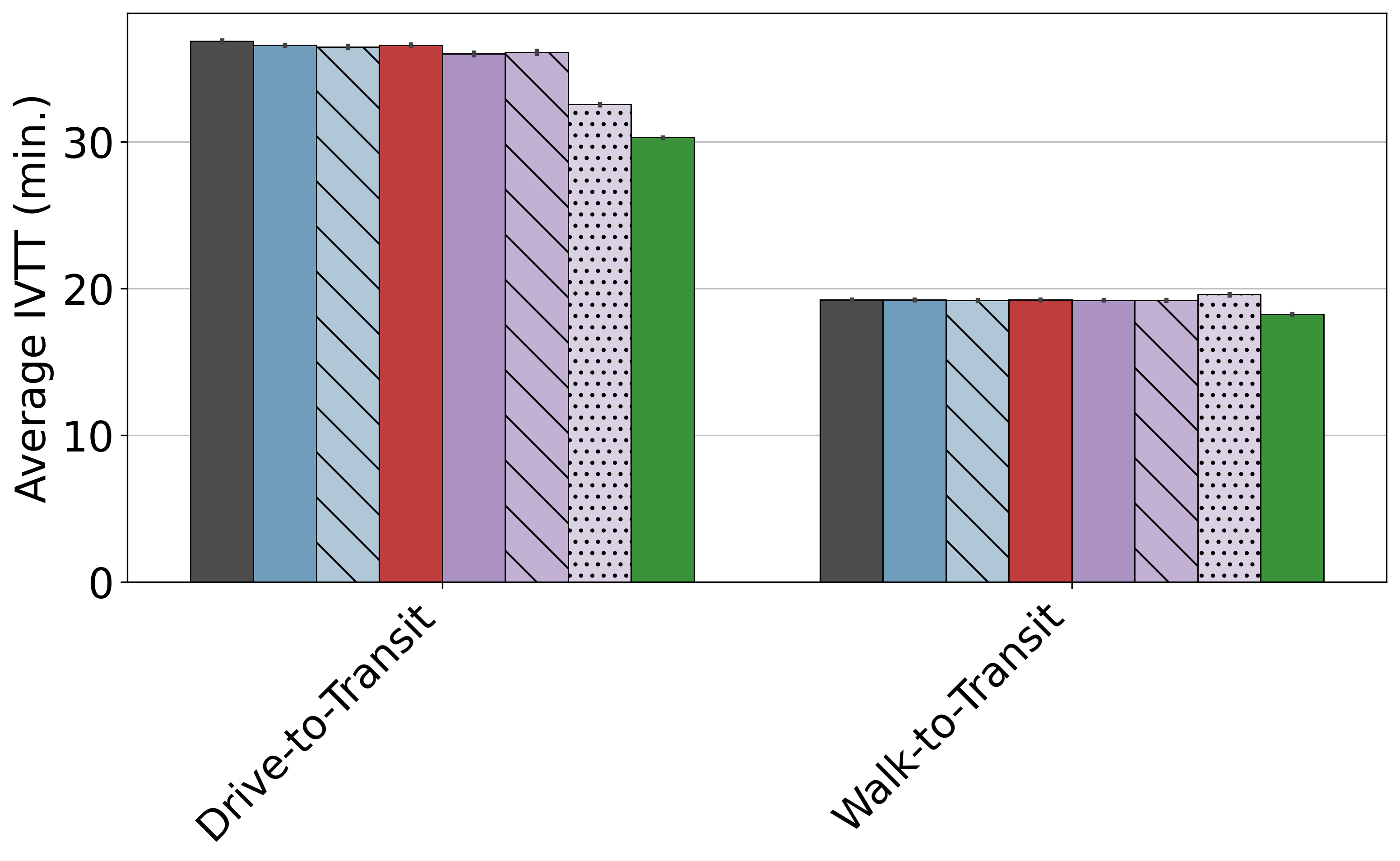}
        \caption{Average IVTT by access mode.}
        \label{fig:transit_avg_ivtt}
    \end{subfigure}
    \hfill
    \begin{subfigure}[t]{0.31\textwidth}
        \centering
        \includegraphics[width=\linewidth]
        {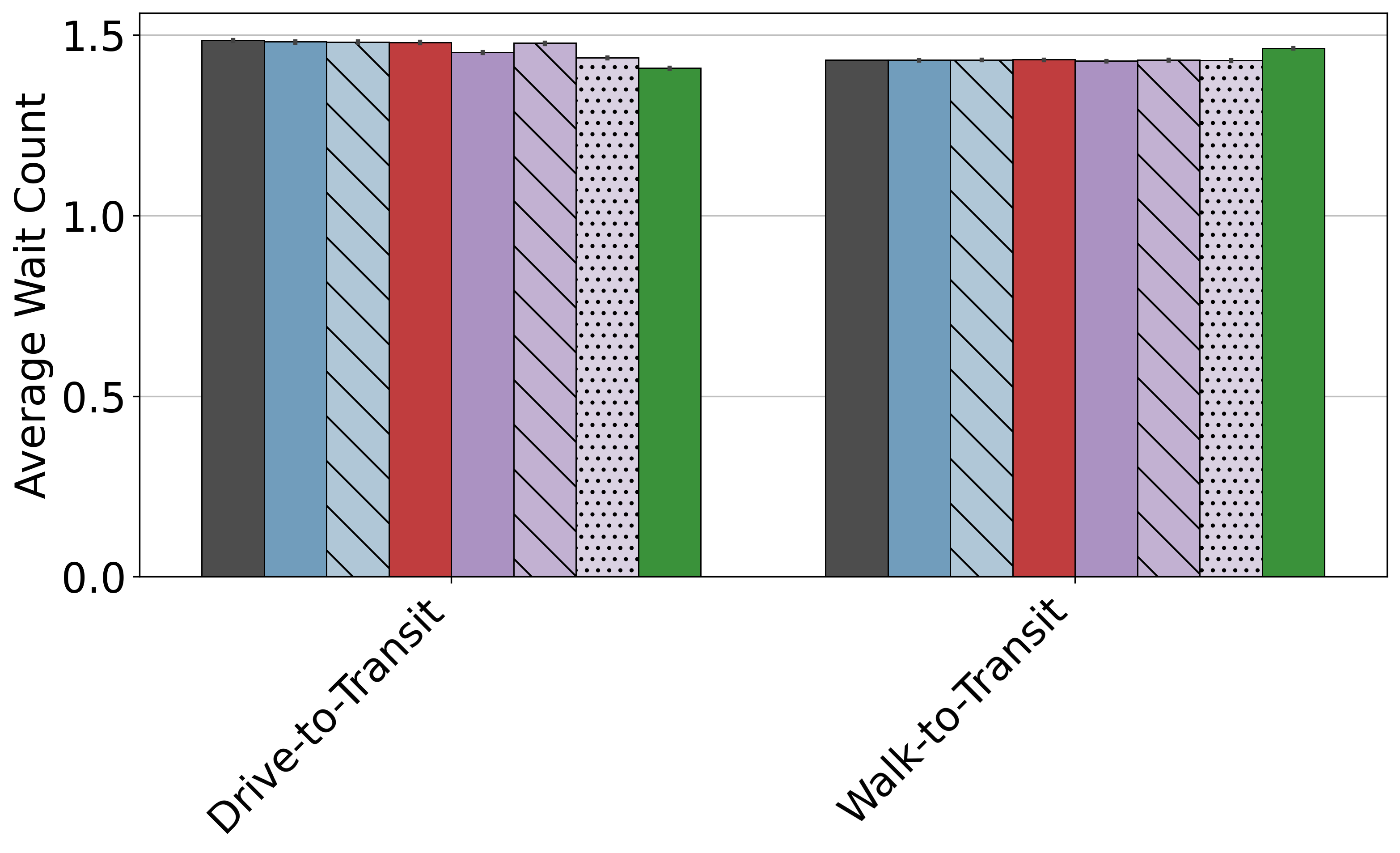}
        \caption{Average wait count by access mode.}
        \label{fig:transit_avg_wait_count}
    \end{subfigure}

    \vspace{0.5em}

    \begin{subfigure}[t]{0.31\textwidth}
        \centering
        \includegraphics[width=\linewidth]
        {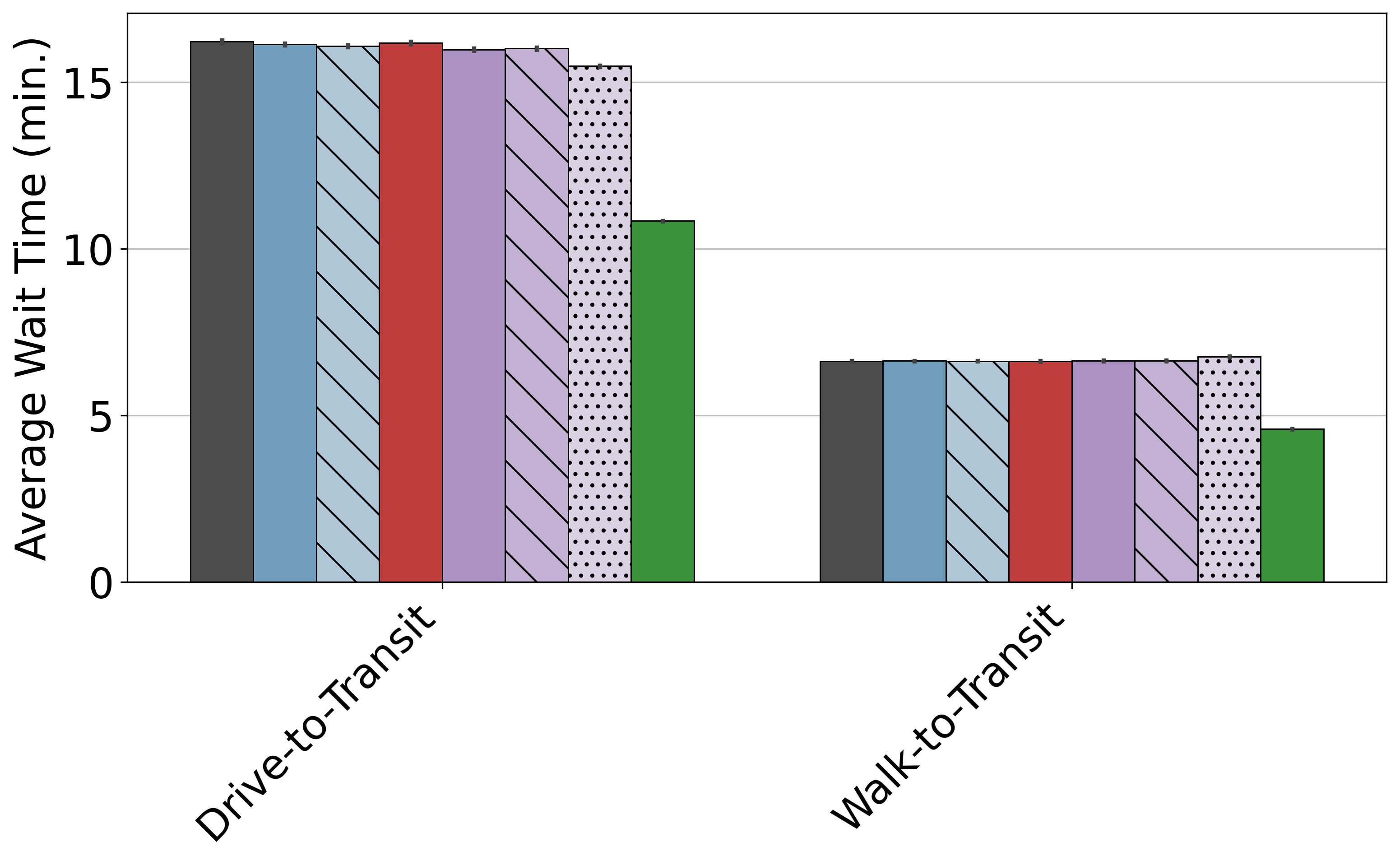}
        \caption{Average wait time by access mode.}
        \label{fig:transit_avg_wait_time}
    \end{subfigure}
    \hfill
    \begin{subfigure}[t]{0.31\textwidth}
        \centering
        \includegraphics[width=\linewidth]
        {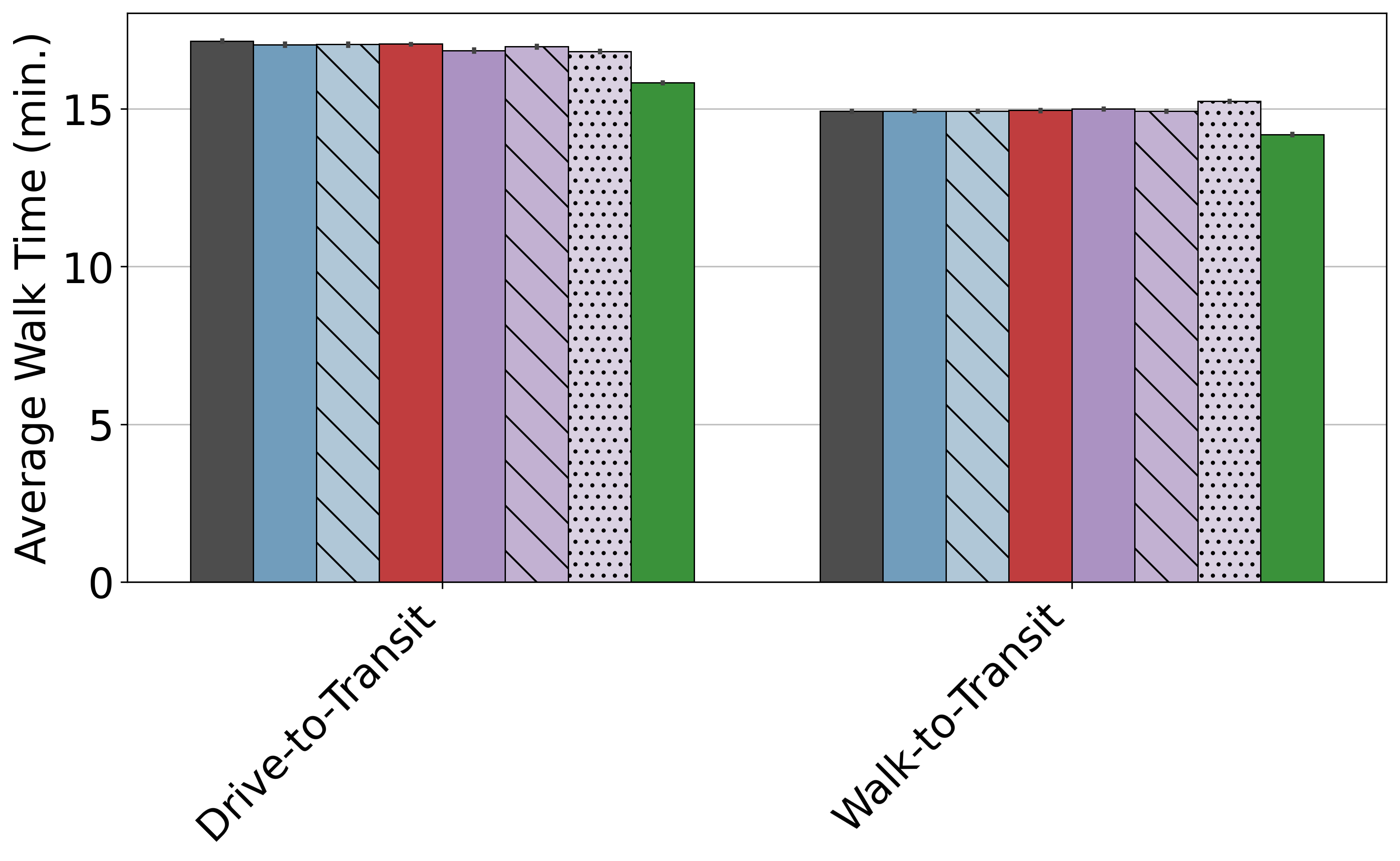}
        \caption{Average walk time by access mode.}
        \label{fig:transit_avg_walk_time}
    \end{subfigure}
    \hfill
    \begin{subfigure}[t]{0.31\textwidth}
        \centering
        \includegraphics[width=\linewidth]
        {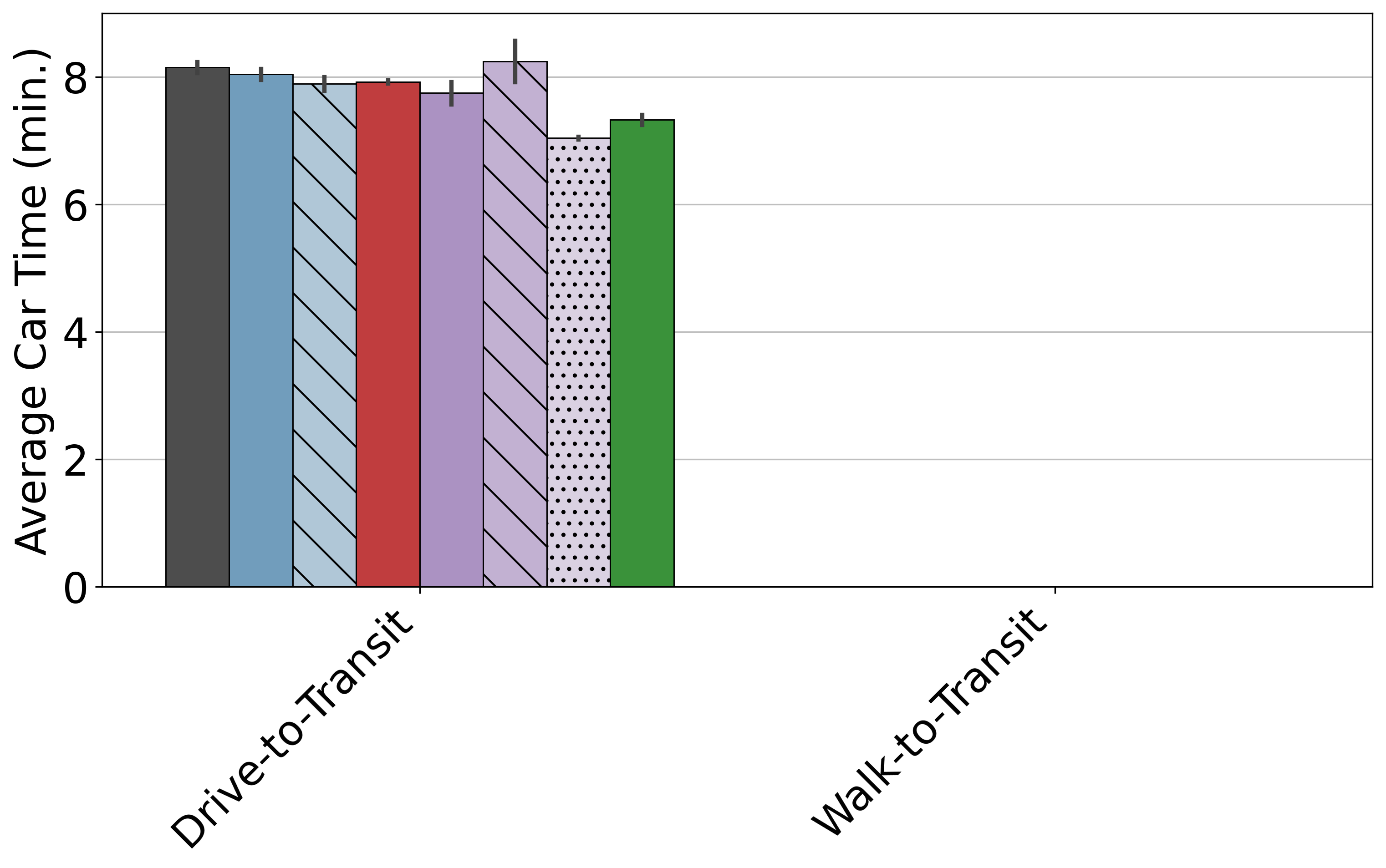}
        \caption{Average car time by access mode.}
        \label{fig:transit_avg_car_time}
    \end{subfigure}

    \caption{Transit experience metrics by access mode across all scenarios.}
    \label{fig:transit_experience_metrics}
\end{figure}

\Cref{fig:transit_experience_metrics} presents the rider-level experience across scenarios and different access modes including drive-to-transit and walk-to-transit. Average wait time shows significant improvement under transit expansion. Drive-to-transit wait time is reduced from approximately 16~min in the BAU to
11~min (reduction of 31\%). Similarly, walk-to-transit decreases from approximately 7~min under BAU to 5~min 
(29\% reduction). 

Average in-vehicle travel time (IVTT) for drive-to-transit trips is decreased from 37~min under BAU to approximately 30~min (19\% reduction). For walk-to-transit, IVTT is slightly reduced (5\%). The significant value of IVTT reduction for drive-to-transit could be due to speed improvement for longer-distance park-and-ride corridors, particularly before riding the rail.

Average car time which is the drive time required to reach a transit station, is reduced approximately 8~min under BAU to approximately 7~min (13\% reduction). Similarly, walk time decreases 6\% for drive-to-transit and 7\% for walk-to-transit compared to BAU. These reductions are consistent with the addition of new stops and BRT corridors that shorten the distance between trip origins and the nearest transit access point.

\subsection{Energy Consumption}
\begin{figure}[!htbp]
    \centering

    \begin{subfigure}[t]{0.47\textwidth}
        \centering
        \includegraphics[width=\linewidth]
        {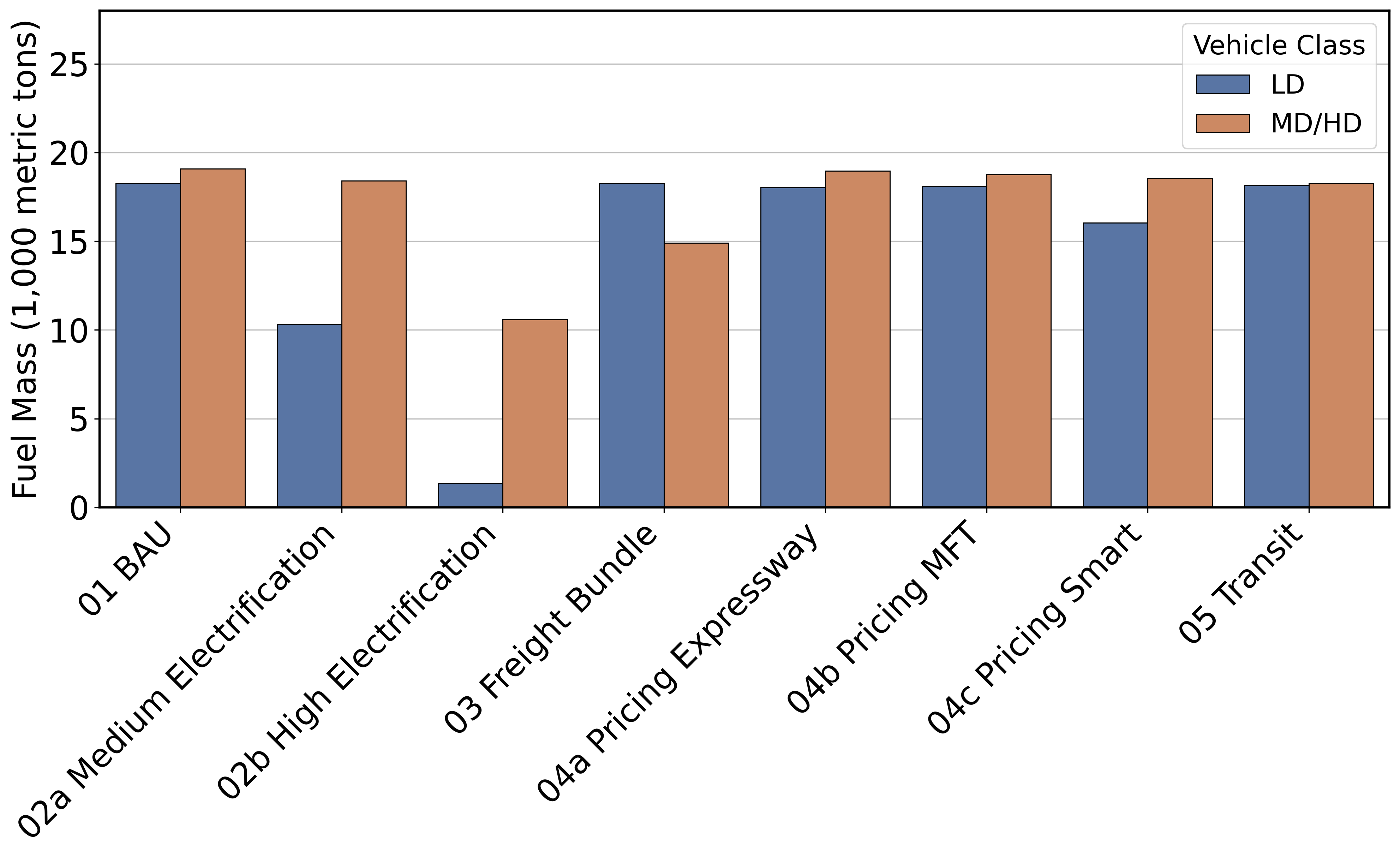}
        \caption{Fuel consumption by vehicle class.}
        \label{fig:emissions_fuel_mass}
    \end{subfigure}
    \hfill
    \begin{subfigure}[t]{0.47\textwidth}
        \centering
        \includegraphics[width=\linewidth]
        {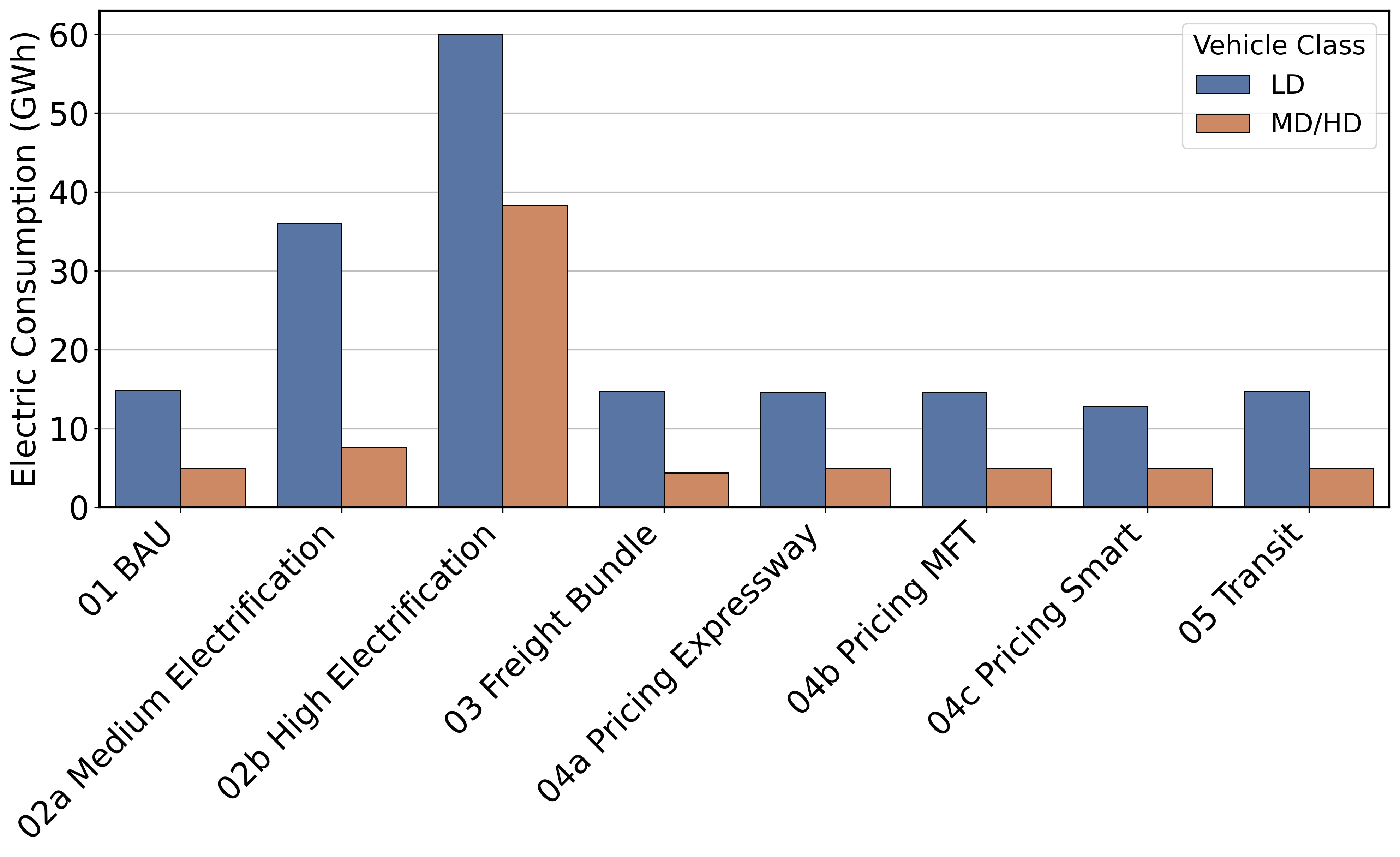}
        \caption{Electric energy consumption by vehicle class.}
        \label{fig:emissions_electric_consumption}
    \end{subfigure}

    \caption{Fuel and electricity consumption across scenarios by LD and MDHD vehicle.}
    \label{fig:emissions_energy_summary}
\end{figure}

\Cref{fig:emissions_energy_summary} presents the fuel (diesel, gasoline, and hydrogen) consumption and electric energy consumption by vehicle class across all scenarios. In the BAU, total fuel consumption reaches 37.5 thousand metric tons (almost equally consumed by LD vehicles and MDHD vehicles, while total electric energy consumption stands at 19.8 GWh (14.8 GWh for LD, and 5 GWh for MDHD).

Electrification scenarios produce the largest reductions in diesel and gasoline mass consumption and the largest increases in electricity demand. Under medium electrification, total fuel mass is decreased by 23.4\% from BAU. The majority of the reduction is driven by LD vehicles (44.0\%) because 55\% of LD vehicles are electrified compared to only 10\% electrified MDHD vehicles. Electric energy consumption rises by 120.1\% in total ( 142.6\% increase by LD, and 53.3\% increase by MDHD). Under high electrification, these effects are intensified substantially. The total fuel mass drops by 68.1\% from BAU, where LD fuel consumption is reduced by 92.6\% because the fleet approaches near-full electrification. MDHD fuel consumption falls by 44.5\%, reflecting the 50\% MDHD EV penetration rate in this scenario. Total electric energy consumption reaches a 395.8\% increase over BAU.

The freight bundle scenario produces an 11.6\% reduction in total fuel mass, mostly driven by MDHD vehicles. Off-hour delivery strategies and long-distance rail diversion from truck reduce MDHD truck fuel consumption by 21.9\%.

Among the pricing scenarios, expressway pricing produces a 10.5\% increase in total fuel mass relative to BAU, mostly driven by LD. This result can be inferred by diversion of LD traffic from tolled expressways to local roads, where lower operating speeds increase per-mile fuel consumption. The mile-based fee and smart pricing scenarios yield modest total reductions of 1.6\% and 7.8\%, respectively. Smart pricing achieves the largest fuel savings among pricing scenarios which reduces LD fuel mass by 12.9\%. The per-mile charge may suppress discretionary auto travel.

\subsection{Electricity Charging}

This study examines three scenarios for electrification analysis: BAU, medium electrification, and high electrification. Although additional scenarios were evaluated, they did not produce significant deviations from the BAU baseline in terms of charging demand or spatial patterns and are therefore excluded from the electrification discussion. The analysis focuses on area types CBD through Exurban, as shown in \Cref{fig:pop_charts}, which account for the vast majority of travel activity and charging demand in the modeled region. The rest of the area presented very small charging activity compared to its land area and was hence excluded from the analysis.

\begin{figure}[!htbp]
  \centering
  \begin{subfigure}[t]{0.32\linewidth}
    \centering
    \includegraphics[width=\linewidth]{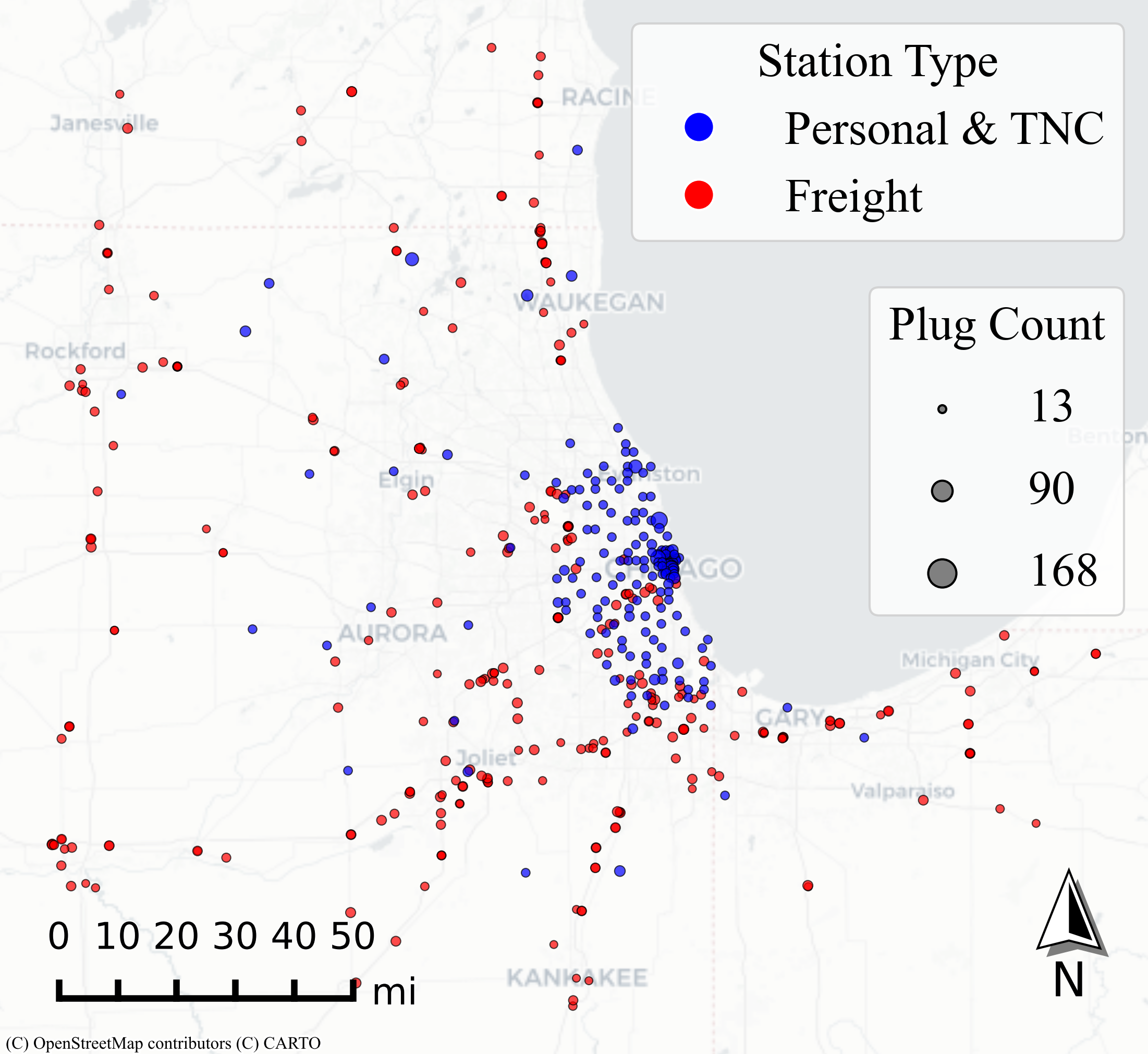}
    \caption{BAU}
    \label{fig:cs_bau}
  \end{subfigure}
  \hfill
  \begin{subfigure}[t]{0.32\linewidth}
    \centering
    \includegraphics[width=\linewidth]{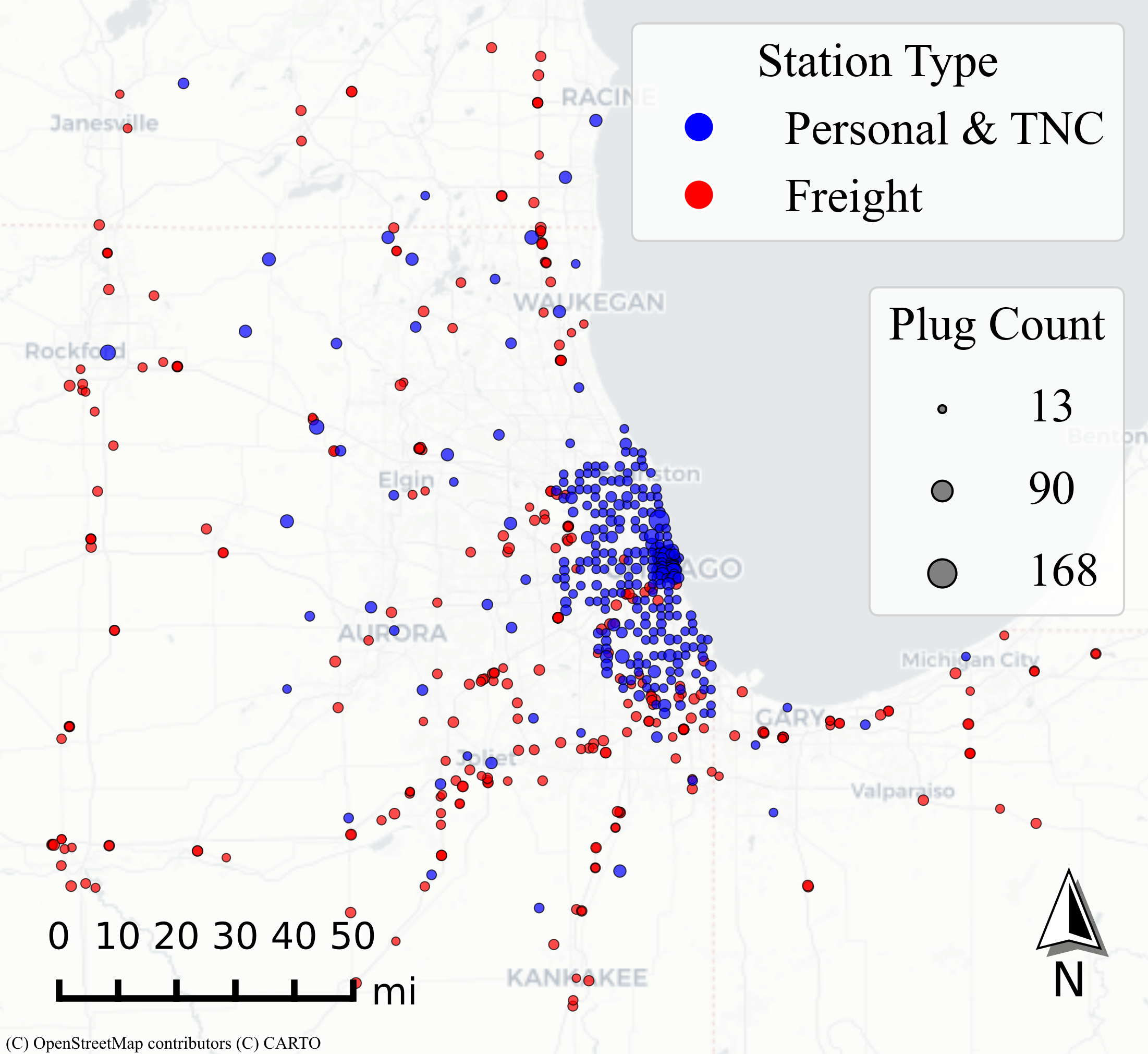}
    \caption{Medium Electrification}
    \label{fig:cs_medium}
  \end{subfigure}
  \hfill
  \begin{subfigure}[t]{0.32\linewidth}
    \centering
    \includegraphics[width=\linewidth]{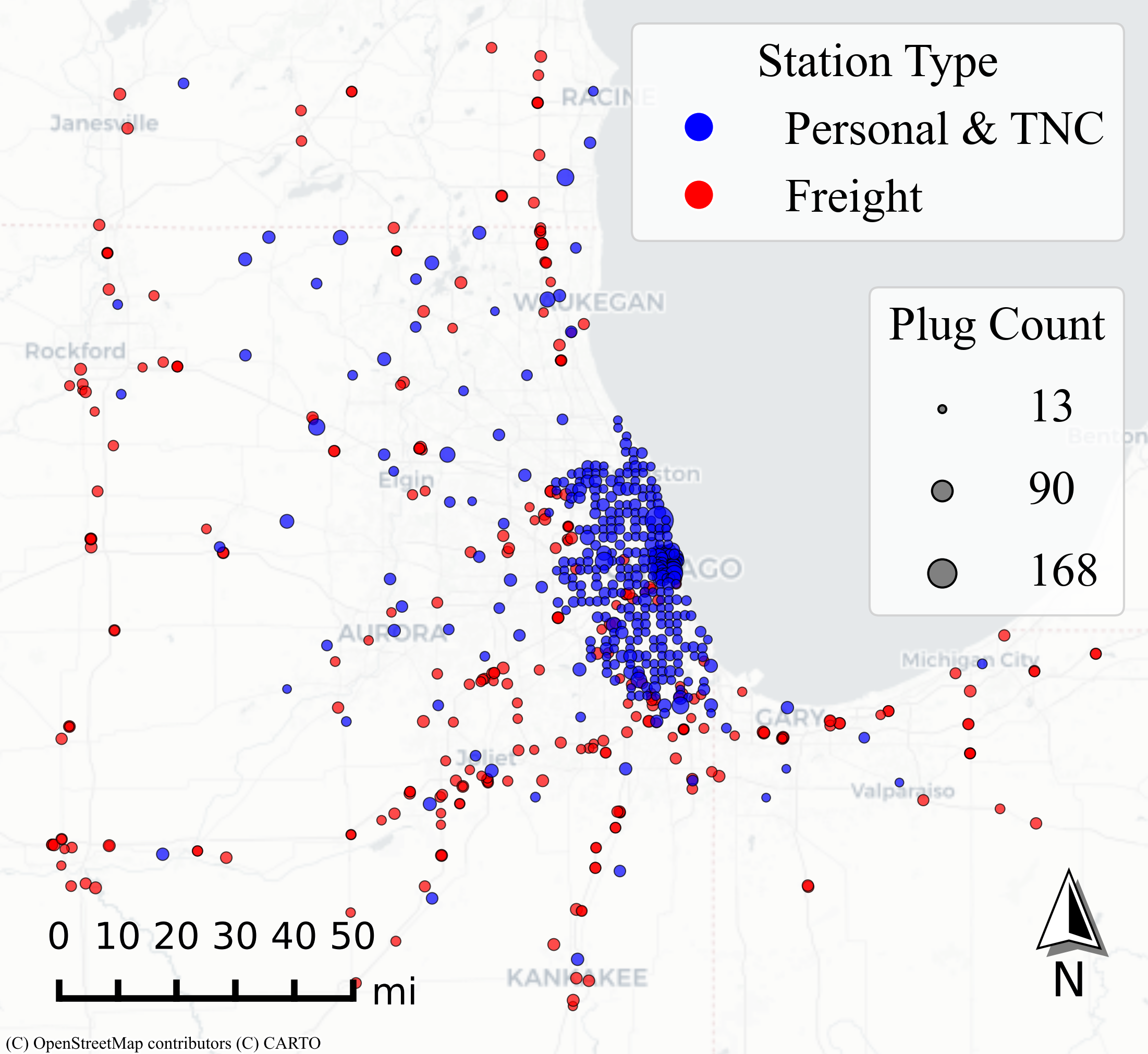}
    \caption{High Electrification}
    \label{fig:cs_high}
  \end{subfigure}
  \caption{EV charging station locations across scenarios. Station count and geographic coverage increase with electrification level.}
  \label{fig:charging_stations}
\end{figure}

\Cref{fig:charging_stations} presents the spatial distribution of EV charging stations across the three scenarios. The BAU scenario provides a baseline network of personal and freight stations concentrated in the urban core. As electrification increases, both the number of stations and the total plug count grow substantially. The medium electrification scenario expands coverage into suburban zones, while the high electrification scenario results in a dense network with widespread geographic reach. Freight charging infrastructure (shown in red) expands primarily at depot locations in industrial and suburban corridors.

\Cref{tab:charging_summary} summarizes charging activity by fleet type, station type, and scenario. Total charging energy consumption increases by approximately 4--8$\times$ from BAU to high electrification depending on the fleet--station combination. Home charging dominates the personal vehicle segment: BAU records approximately 6.1~GWh across 897,000 home events, rising to 16.3~GWh (2.4 million events) in Medium and 27.7~GWh (4.1 million events) in High. Public personal charging grows from 1.1~GWh to 4.8~GWh over the same range. On the freight side, depot charging energy consumption increases from 2.1~GWh (14,500 events) in BAU to 15.9~GWh (105,000 events) in High, while dock charging grows from 0.27~GWh to 2.1~GWh. TNC charging emerges only in the Medium and High scenarios with modest volumes. Average per-session energy consumption remains relatively stable across scenarios, indicating that load growth is driven predominantly by fleet size rather than individual charging behavior.

\begin{table}[!htb]
\centering
\caption{Charging summary by fleet type, station type, and scenario.}
\label{tab:charging_summary}
\footnotesize
\begin{tabular}{llrrrrr}
\toprule
\textbf{Scenario} & \textbf{Fleet -- Station} & \textbf{Events} & \textbf{Energy (GWh)} & \textbf{Avg kWh} & \textbf{Avg Min} \\
\midrule
\multirow{4}{*}{BAU}
  & Personal -- Home   & 896,604  & 6.06  & 6.76  & 56.3 \\
  & Personal -- Public & 39,384   & 1.07  & 27.13 & 10.8 \\
  & Freight -- Depot   & 14,484   & 2.08  & 143.5 & 24.6 \\
  & Freight -- Dock    & 14,896   & 0.27  & 18.3  & 57.3 \\
\midrule
\multirow{5}{*}{\shortstack[l]{Medium\\Elec.}}
  & Personal -- Home   & 2,410,048 & 16.28 & 6.76  & 56.3 \\
  & Personal -- Public & 102,180   & 2.90  & 28.36 & 11.4 \\
  & Freight -- Depot   & 25,020    & 3.28  & 131.2 & 22.5 \\
  & Freight -- Dock    & 23,268    & 0.43  & 18.4  & 57.4 \\
  & TNC -- Public      & 60        & 0.006 & 93.5  & 430.6 \\
\midrule
\multirow{5}{*}{\shortstack[l]{High\\Elec.}}
  & Personal -- Home   & 4,106,920 & 27.66 & 6.73  & 56.1 \\
  & Personal -- Public & 167,196   & 4.78  & 28.62 & 11.5 \\
  & Freight -- Depot   & 105,000   & 15.90 & 151.4 & 25.9 \\
  & Freight -- Dock    & 115,468   & 2.12  & 18.4  & 57.5 \\
  & TNC -- Public      & 152       & 0.013 & 85.3  & 35.8 \\
\bottomrule
\end{tabular}
\end{table}

\begin{figure}[ht]
  \centering
  \begin{subfigure}[t]{0.32\linewidth}
    \centering
    \includegraphics[width=\linewidth]{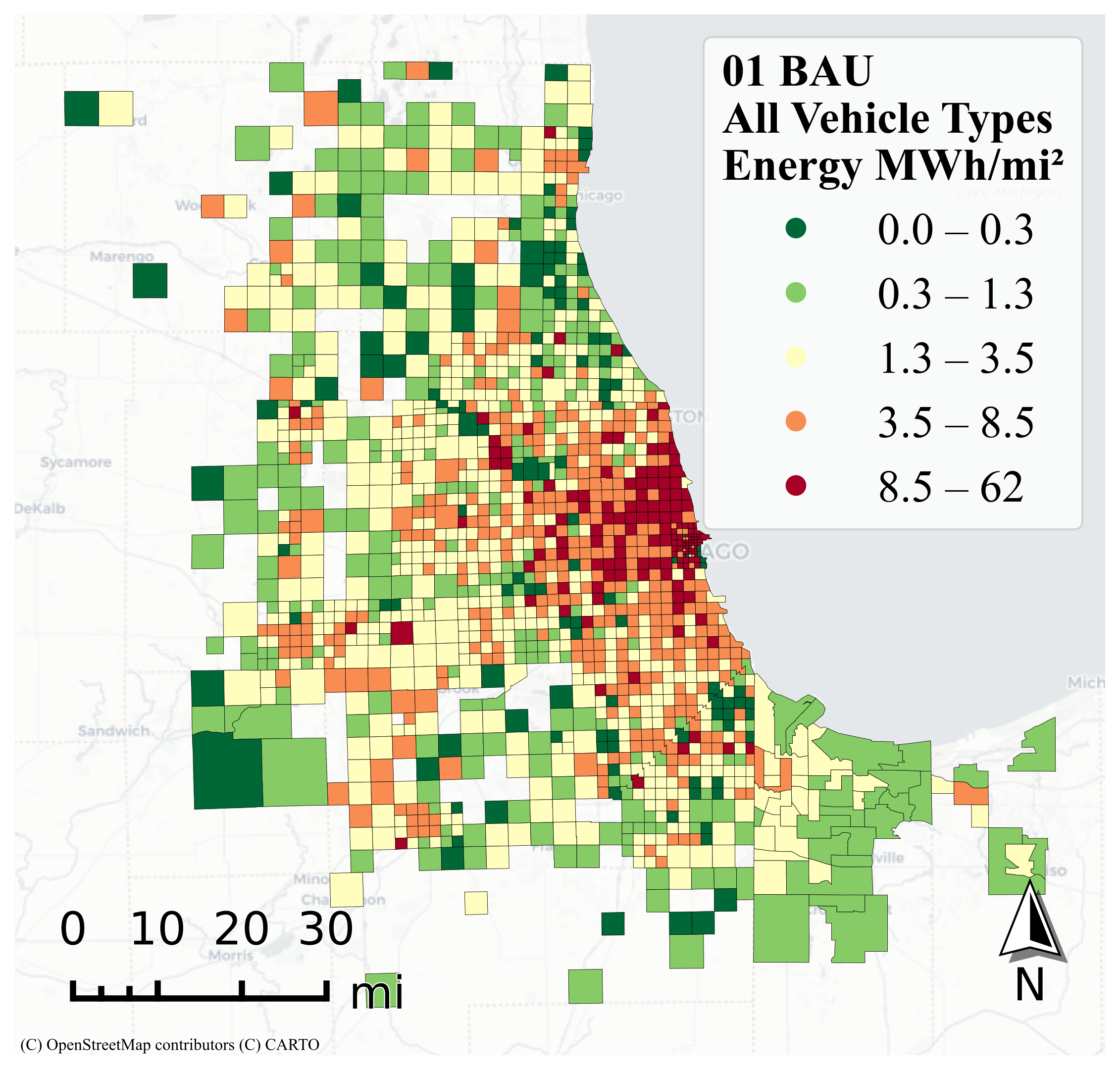}
    \caption{BAU}
    \label{fig:map_bau}
  \end{subfigure}
  \hfill
  \begin{subfigure}[t]{0.32\linewidth}
    \centering
    \includegraphics[width=\linewidth]{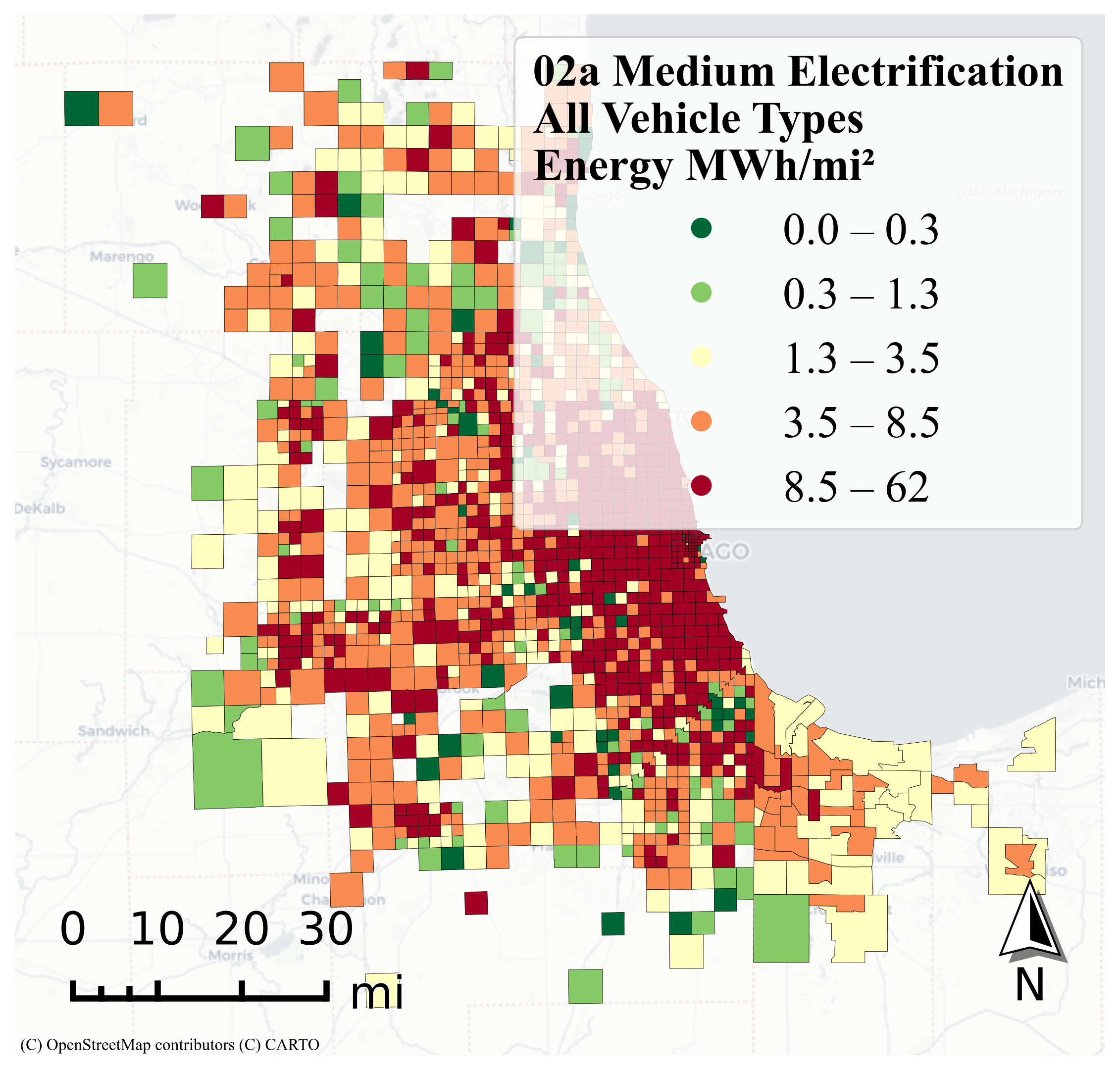}
    \caption{Medium Electrification}
    \label{fig:map_medium}
  \end{subfigure}
  \hfill
  \begin{subfigure}[t]{0.32\linewidth}
    \centering
    \includegraphics[width=\linewidth]{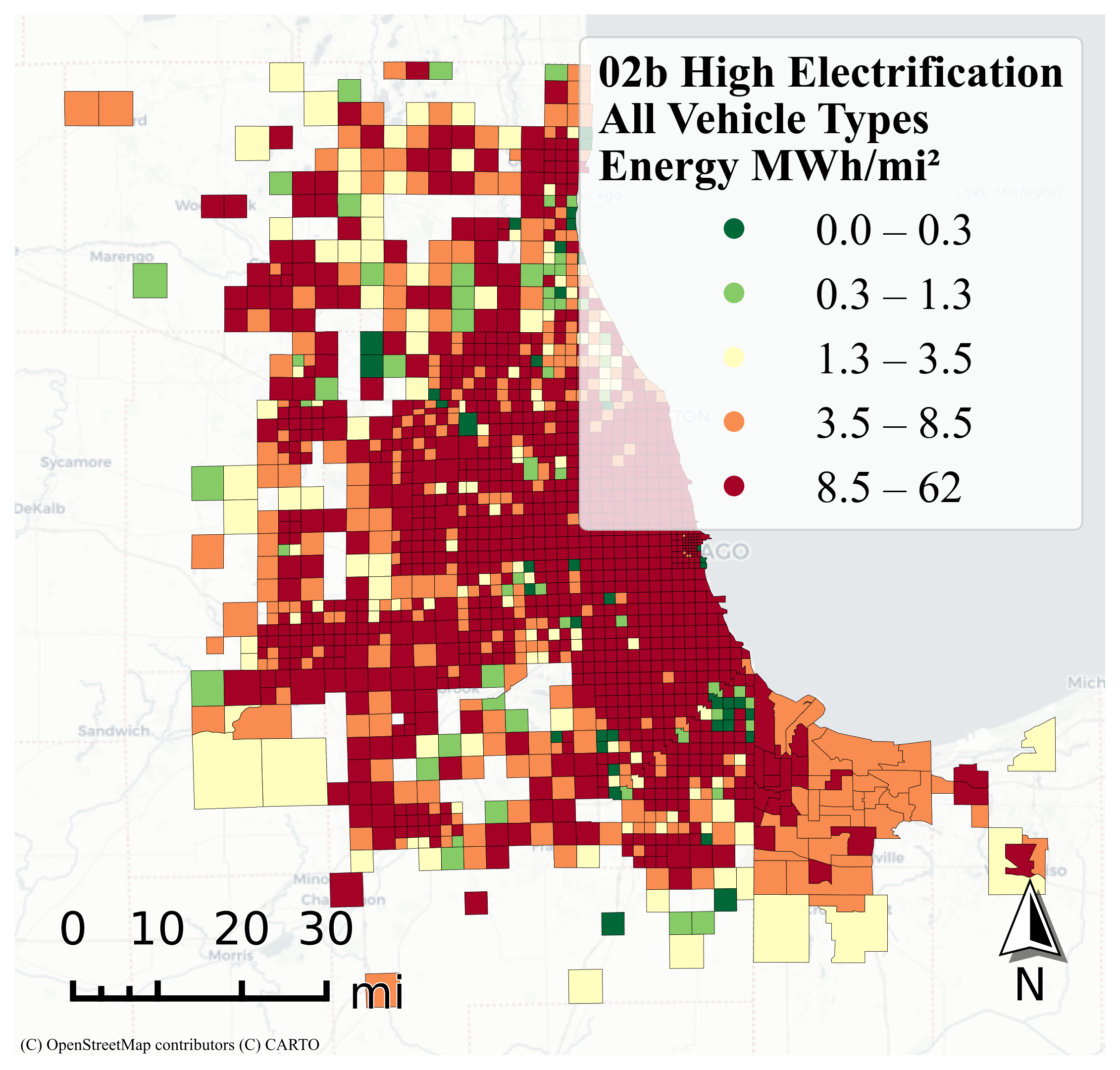}
    \caption{High Electrification}
    \label{fig:map_high}
  \end{subfigure}
  \caption{Spatial distribution of total charging energy density (MWh/mi\textsuperscript{2}) across scenarios. Energy maps cover area types CBD--Exurban only (see \Cref{fig:pop_charts}).}
  \label{fig:energy_maps}
\end{figure}

\Cref{fig:energy_maps} presents the spatial distribution of total charging energy density (all vehicle types) across the three scenarios. Charging energy density concentrates heavily in the CBD and inner urban zones. The medium electrification scenario introduces a visible expansion of mid-density charging zones into suburban areas, while the high electrification scenario produces a pronounced high-density cluster along the lakefront and CBD, reflecting the compounding effect of high EV penetration and dense trip generation in the urban core. This spatial concentration of charging demand in high-activity urban zones could place localized stress on distribution grid infrastructure. Energy and power maps are computed for area types CBD through Exurban only, as shown in \Cref{fig:pop_charts}.

\begin{figure}[!htbp]
  \centering
  \begin{subfigure}[c]{0.35\linewidth}\centering\includegraphics[width=\linewidth]{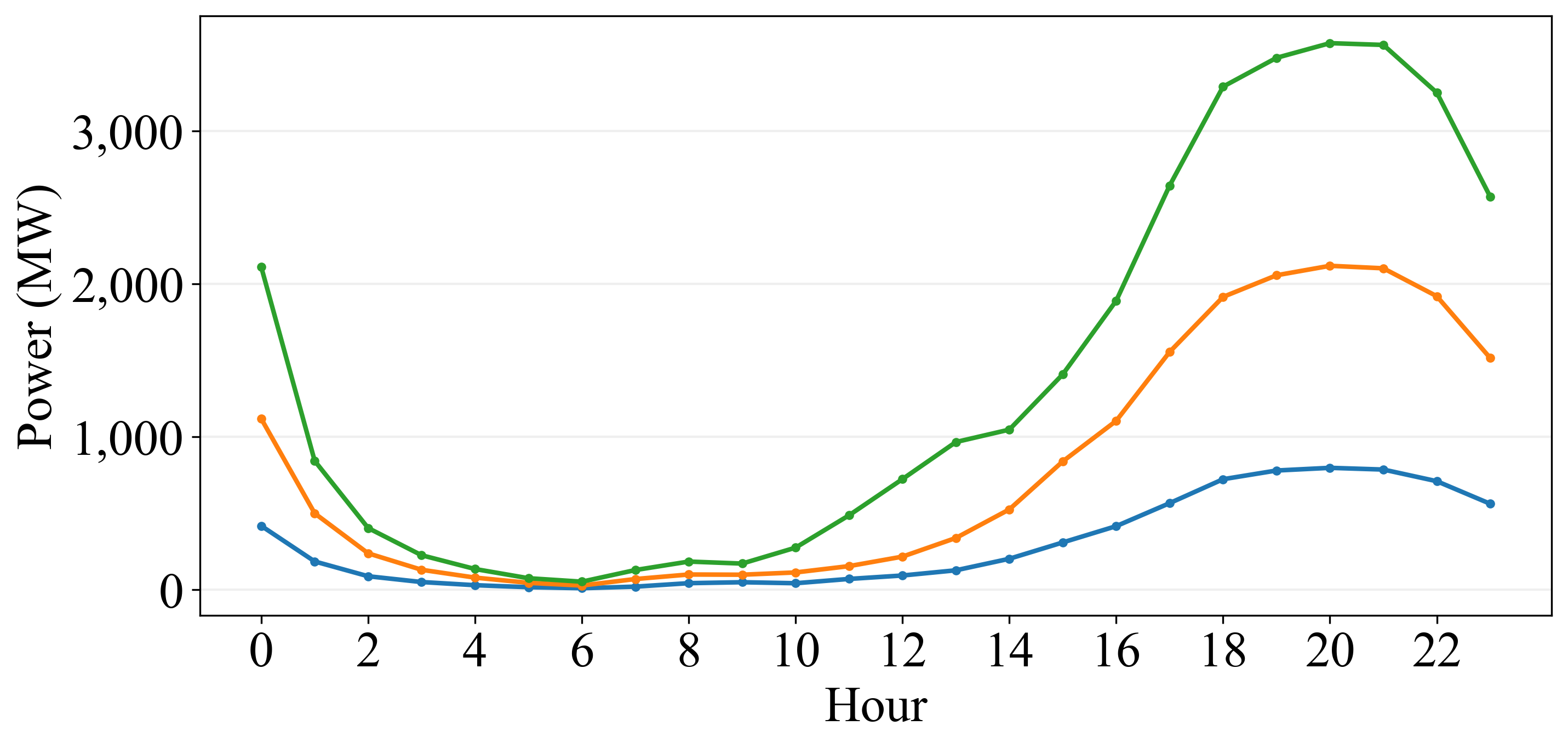}\caption{All Vehicles -- Power}\end{subfigure}\quad
  \begin{subfigure}[c]{0.35\linewidth}\centering\includegraphics[width=\linewidth]{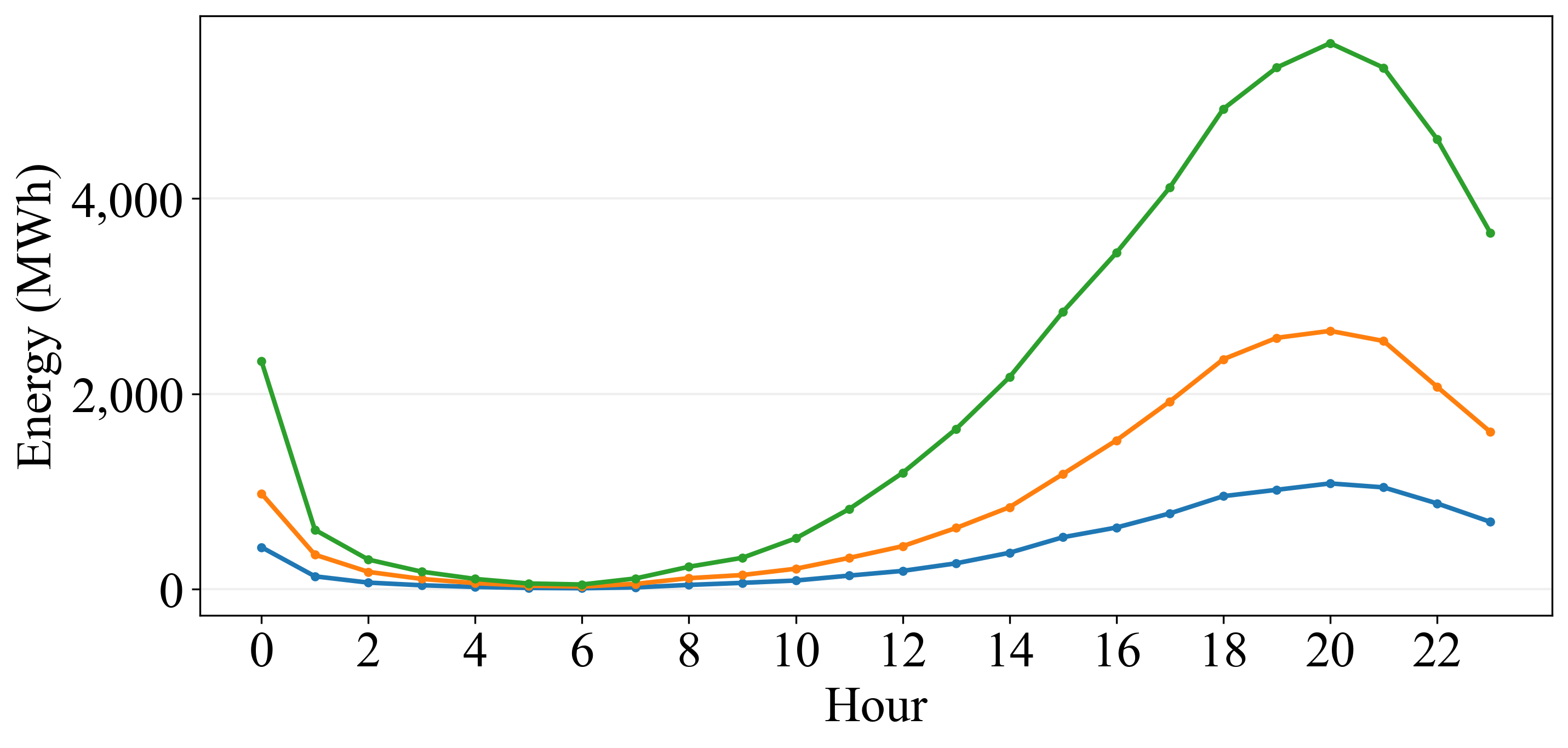}\caption{All Vehicles -- Energy}\end{subfigure}\\
  \begin{subfigure}[c]{0.35\linewidth}\centering\includegraphics[width=\linewidth]{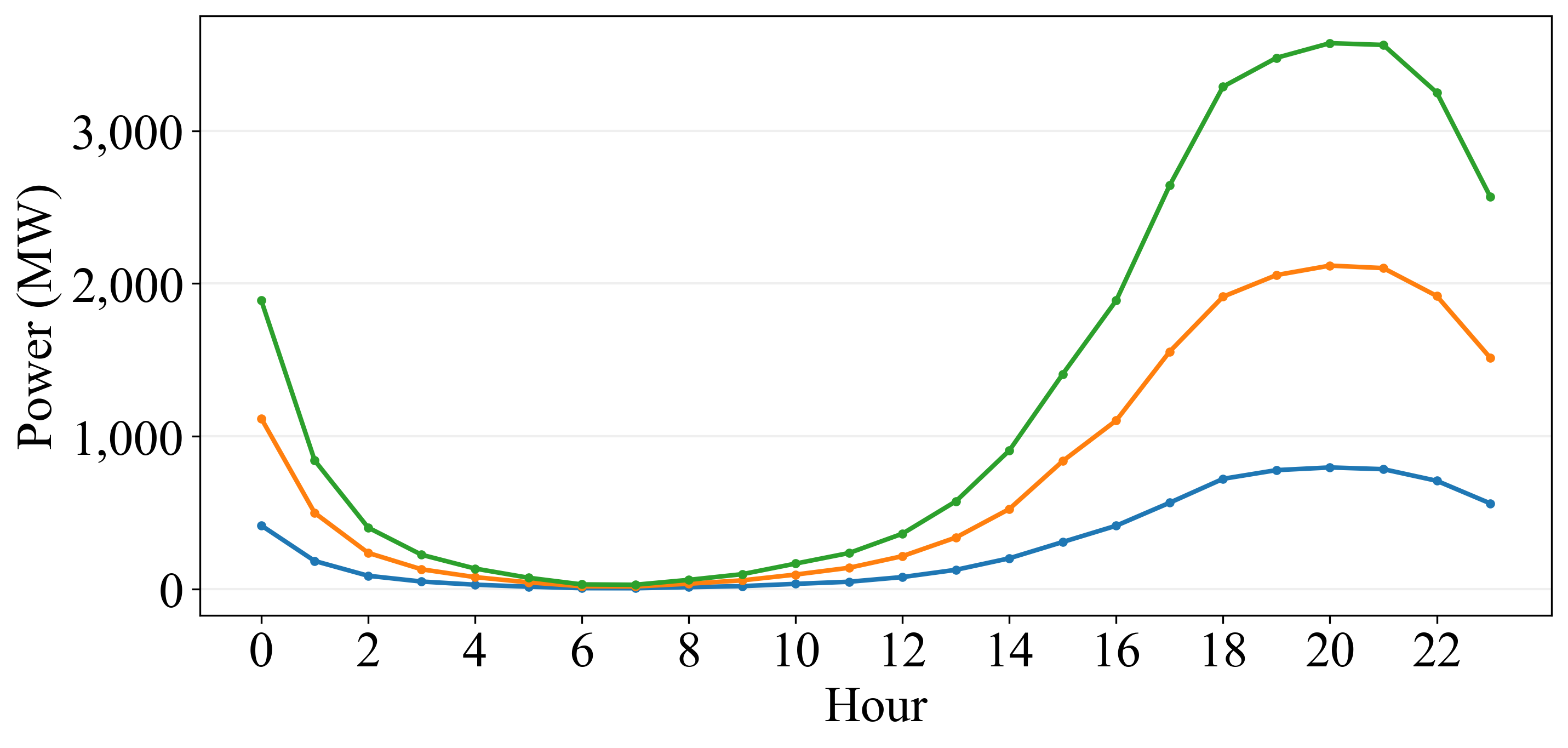}\caption{Personal Home -- Power}\end{subfigure}\quad
  \begin{subfigure}[c]{0.35\linewidth}\centering\includegraphics[width=\linewidth]{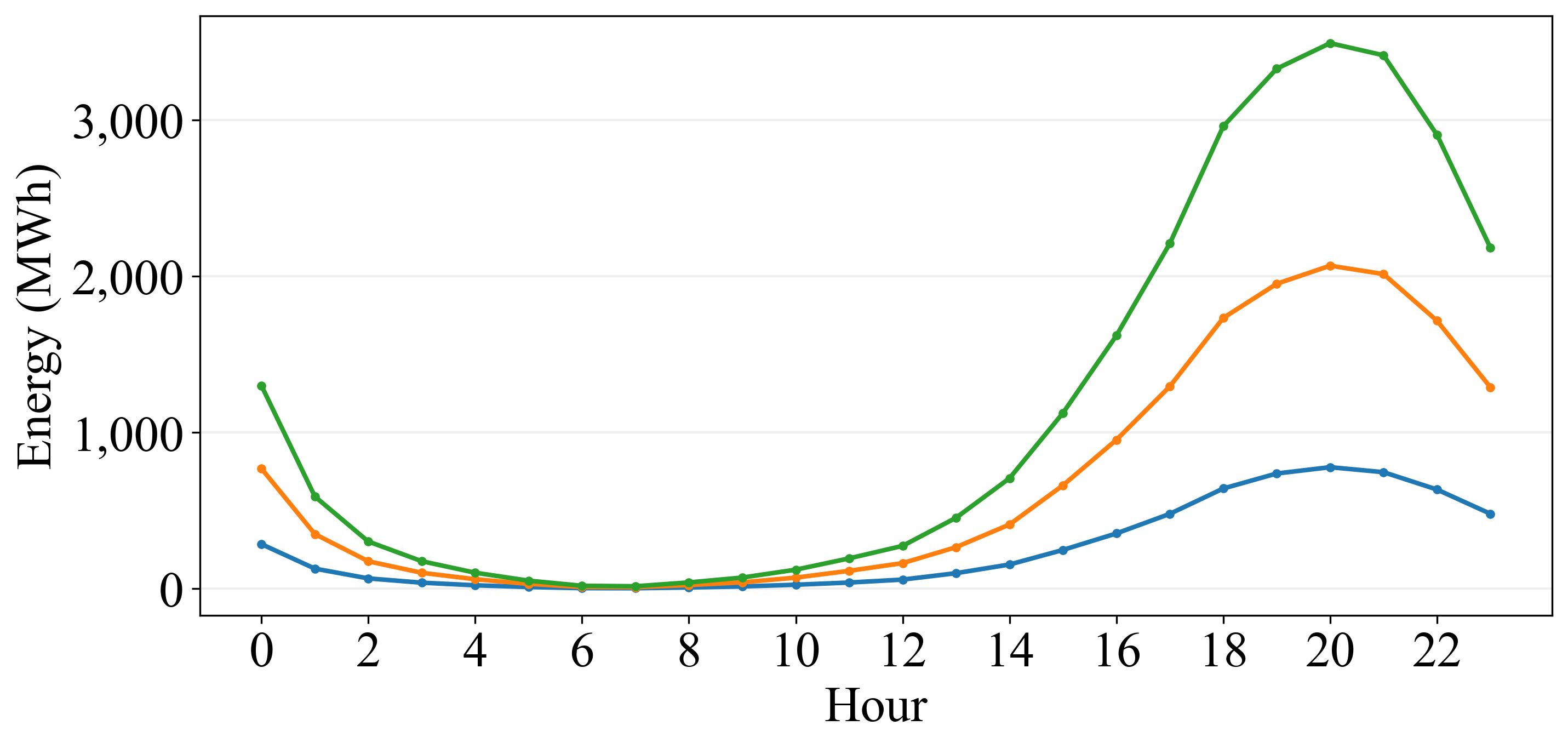}\caption{Personal Home -- Energy}\end{subfigure}\\
  \begin{subfigure}[c]{0.35\linewidth}\centering\includegraphics[width=\linewidth]{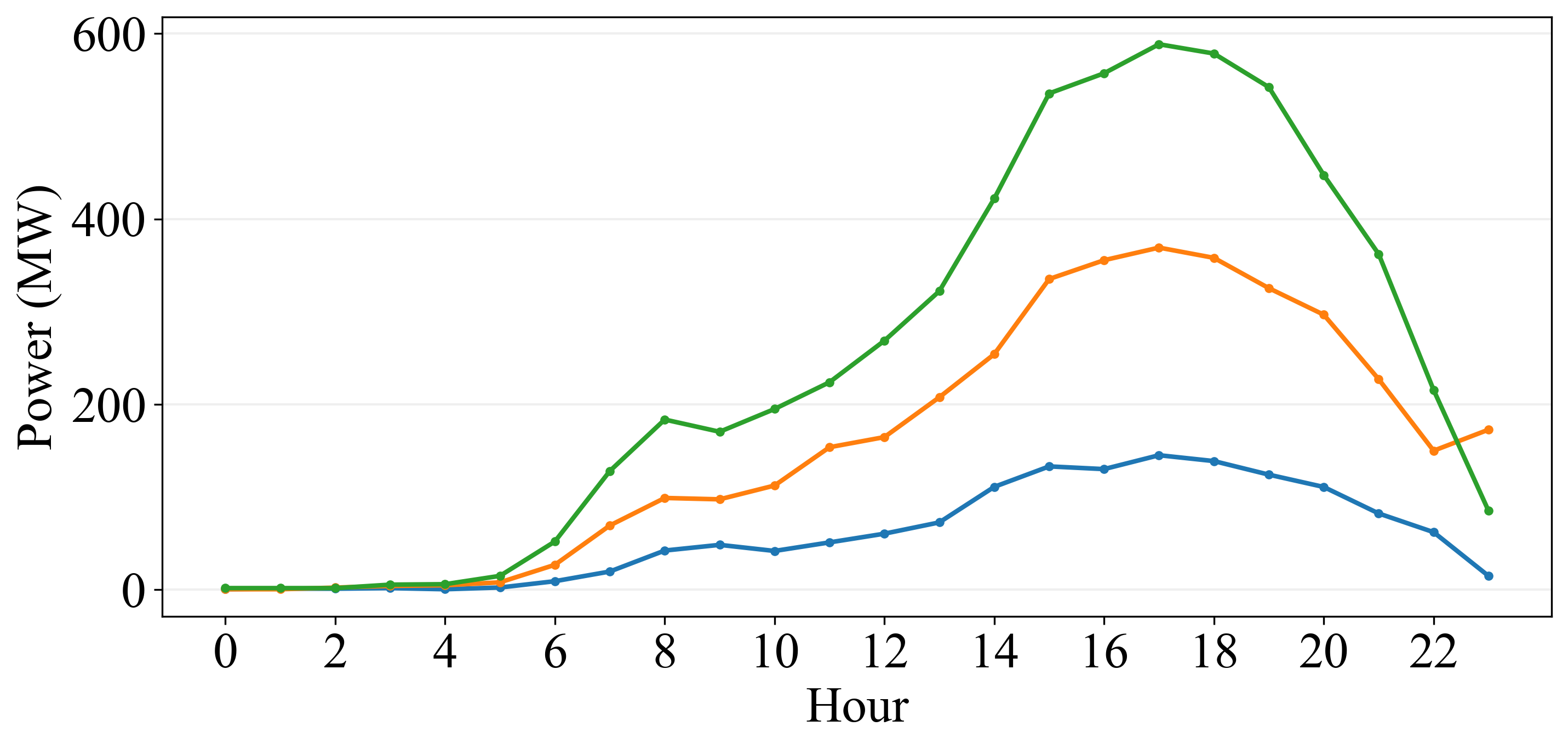}\caption{Personal Public -- Power}\end{subfigure}\quad
  \begin{subfigure}[c]{0.35\linewidth}\centering\includegraphics[width=\linewidth]{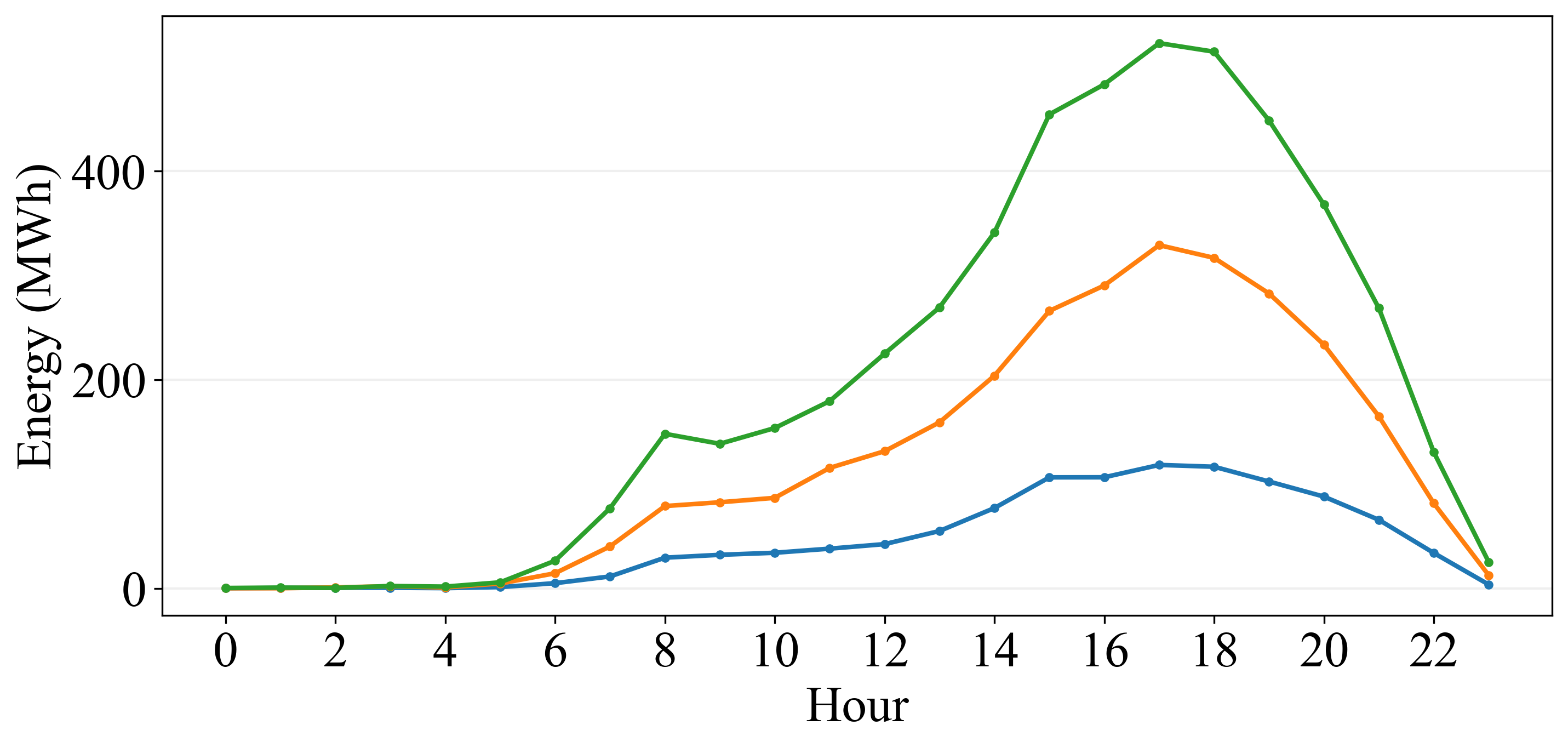}\caption{Personal Public -- Energy}\end{subfigure}\\
  \begin{subfigure}[c]{0.35\linewidth}\centering\includegraphics[width=\linewidth]{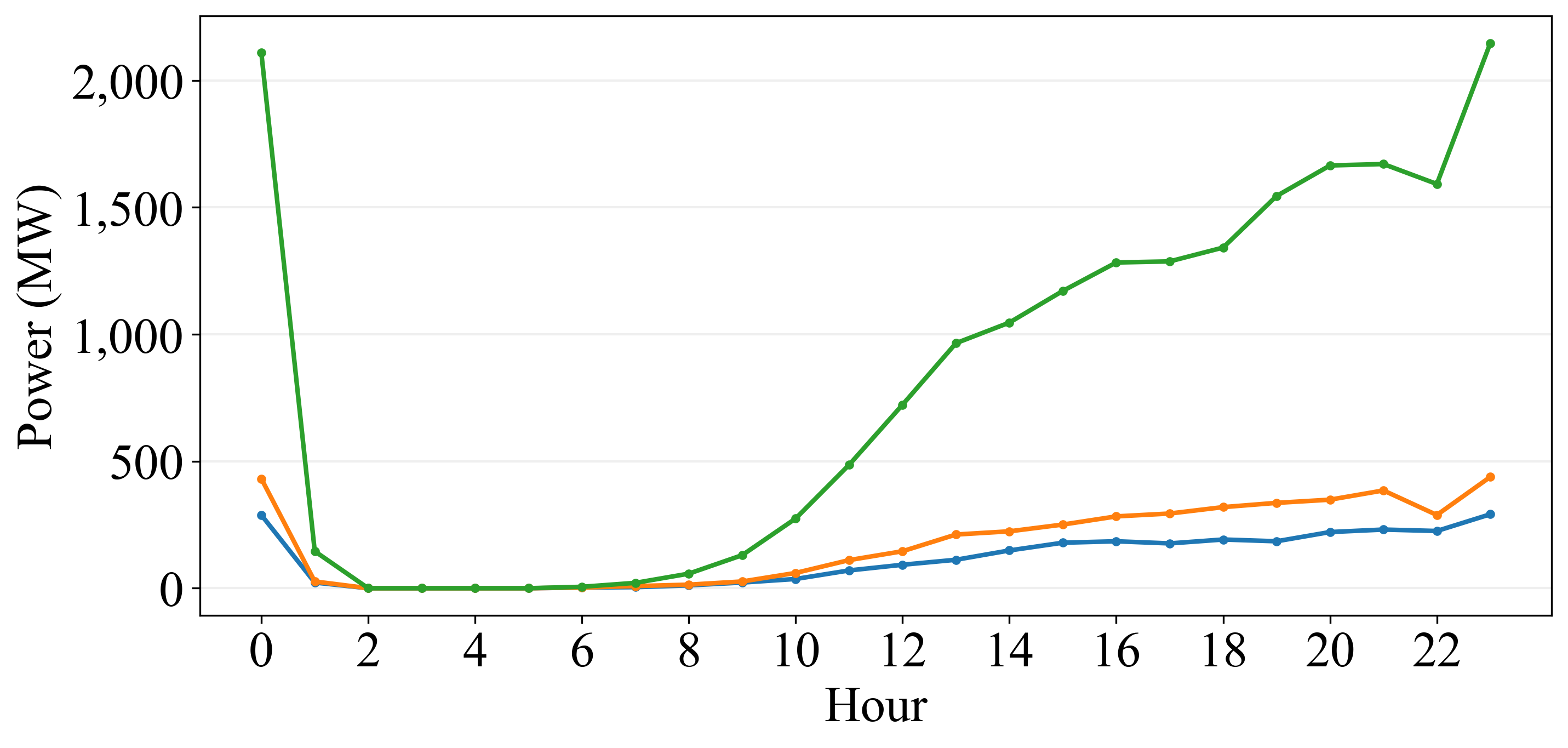}\caption{Freight Depot -- Power}\end{subfigure}\quad
  \begin{subfigure}[c]{0.35\linewidth}\centering\includegraphics[width=\linewidth]{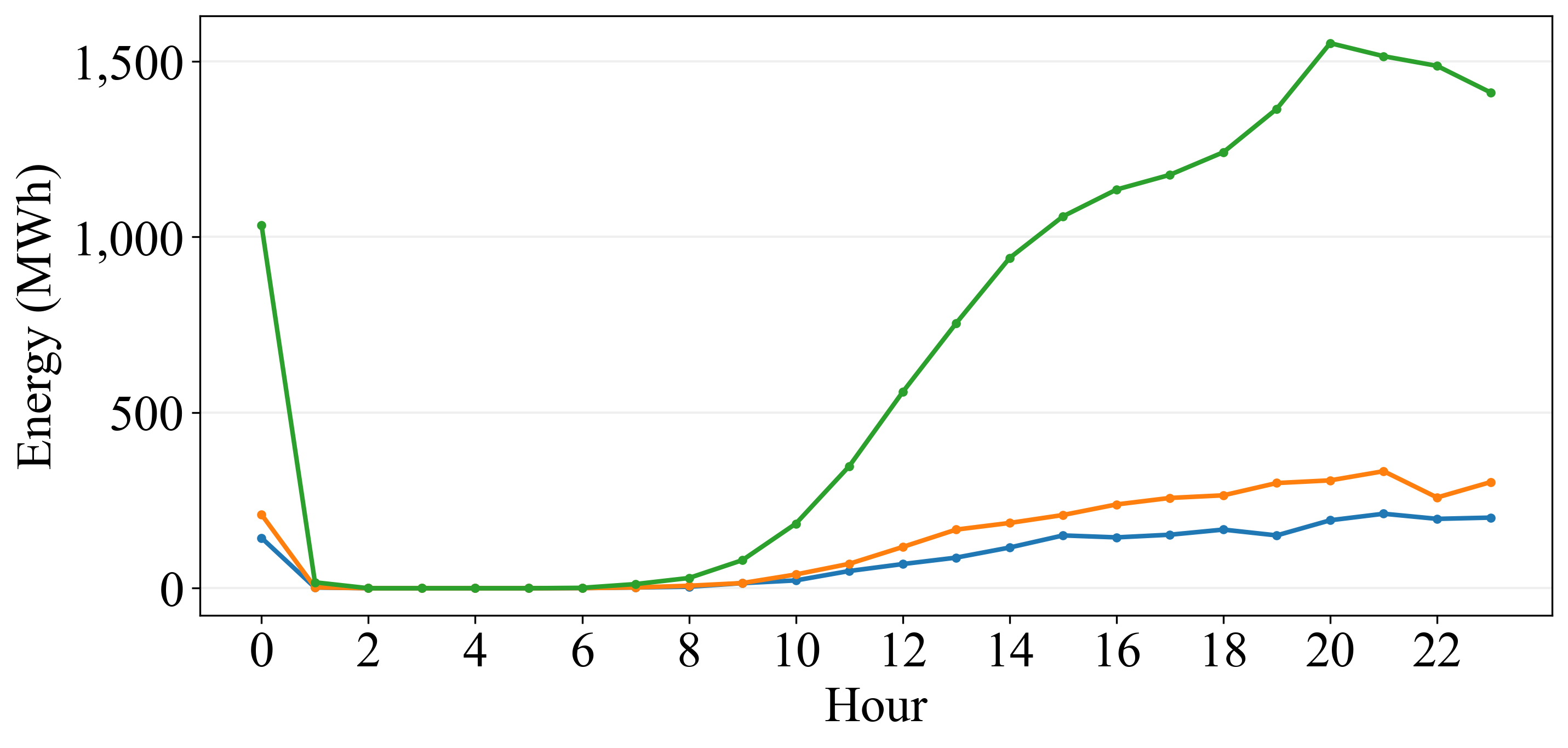}\caption{Freight Depot -- Energy}\end{subfigure}\\
  \begin{subfigure}[c]{0.35\linewidth}\centering\includegraphics[width=\linewidth]{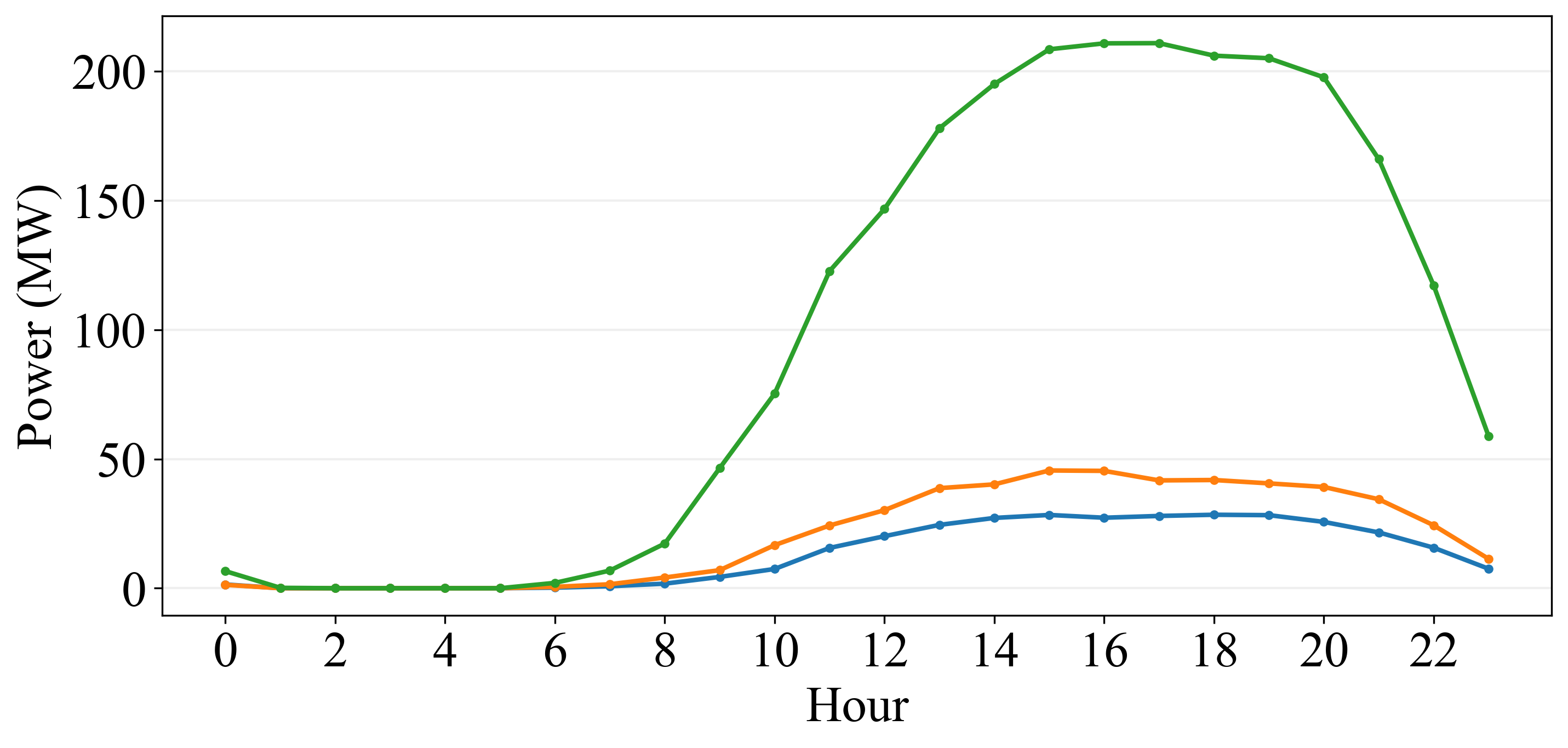}\caption{Freight Dock -- Power}\end{subfigure}\quad
  \begin{subfigure}[c]{0.35\linewidth}\centering\includegraphics[width=\linewidth]{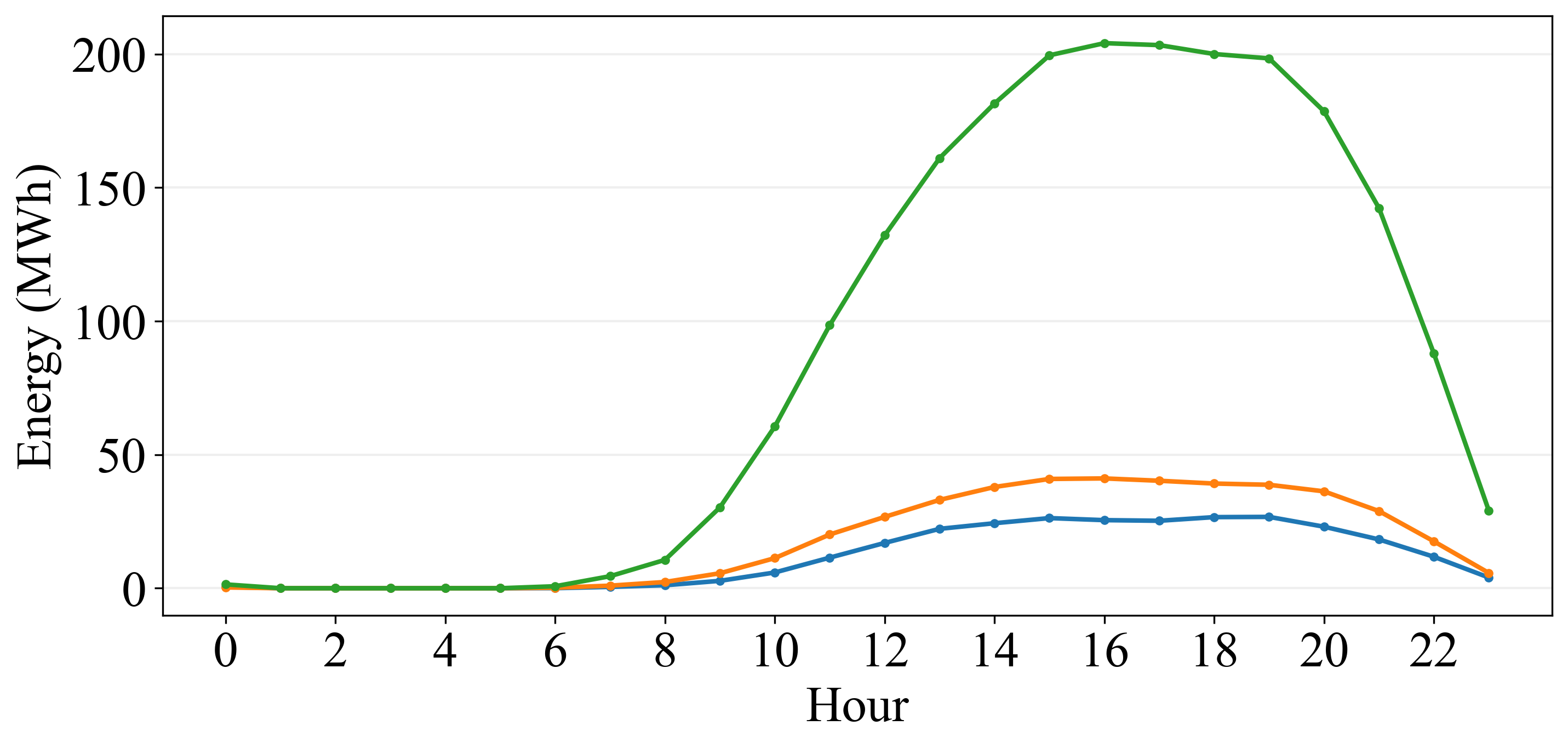}\caption{Freight Dock -- Energy}\end{subfigure}\\
  \begin{subfigure}[c]{0.35\linewidth}\centering\includegraphics[width=\linewidth]{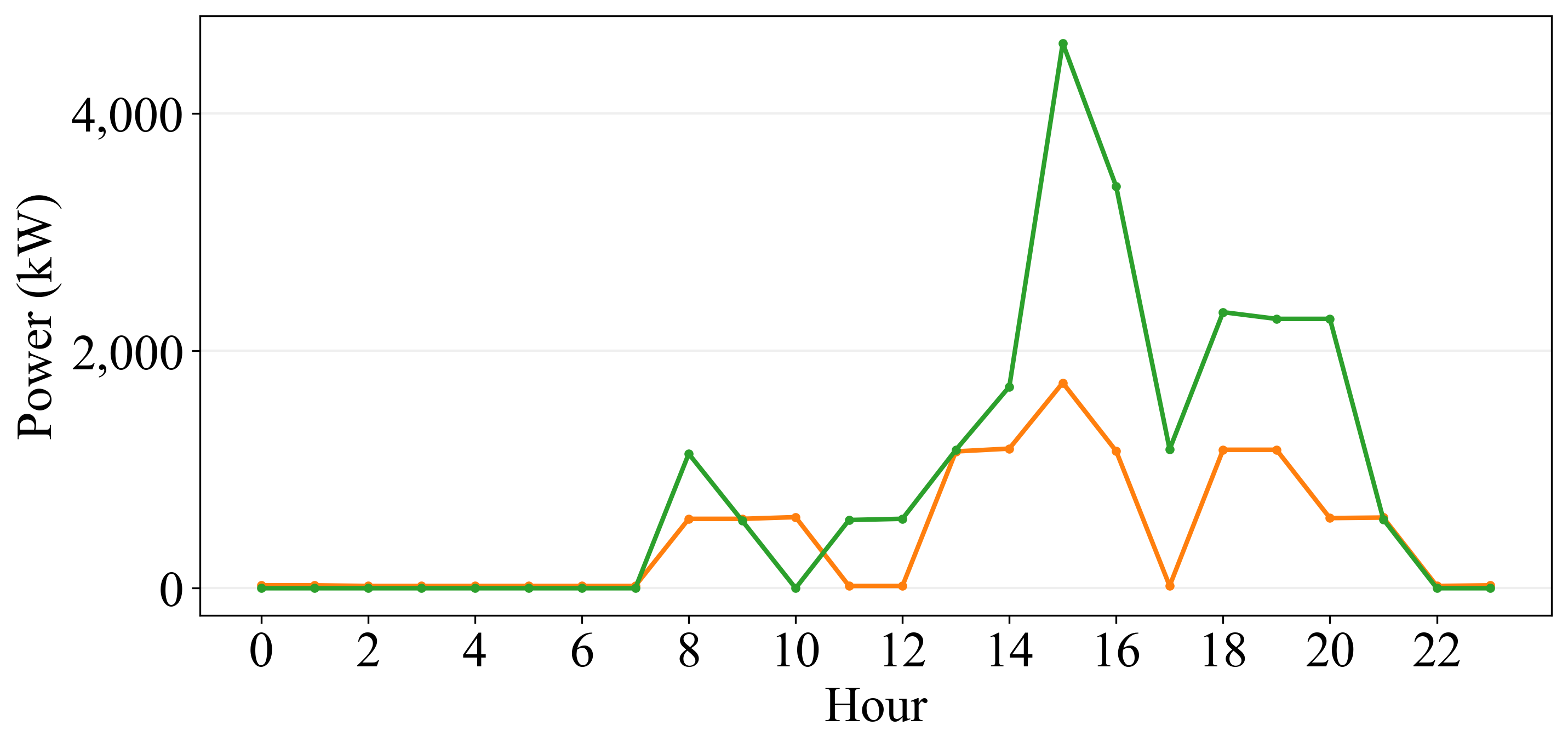}\caption{TNC Public -- Power}\end{subfigure}\quad
  \begin{subfigure}[c]{0.35\linewidth}\centering\includegraphics[width=\linewidth]{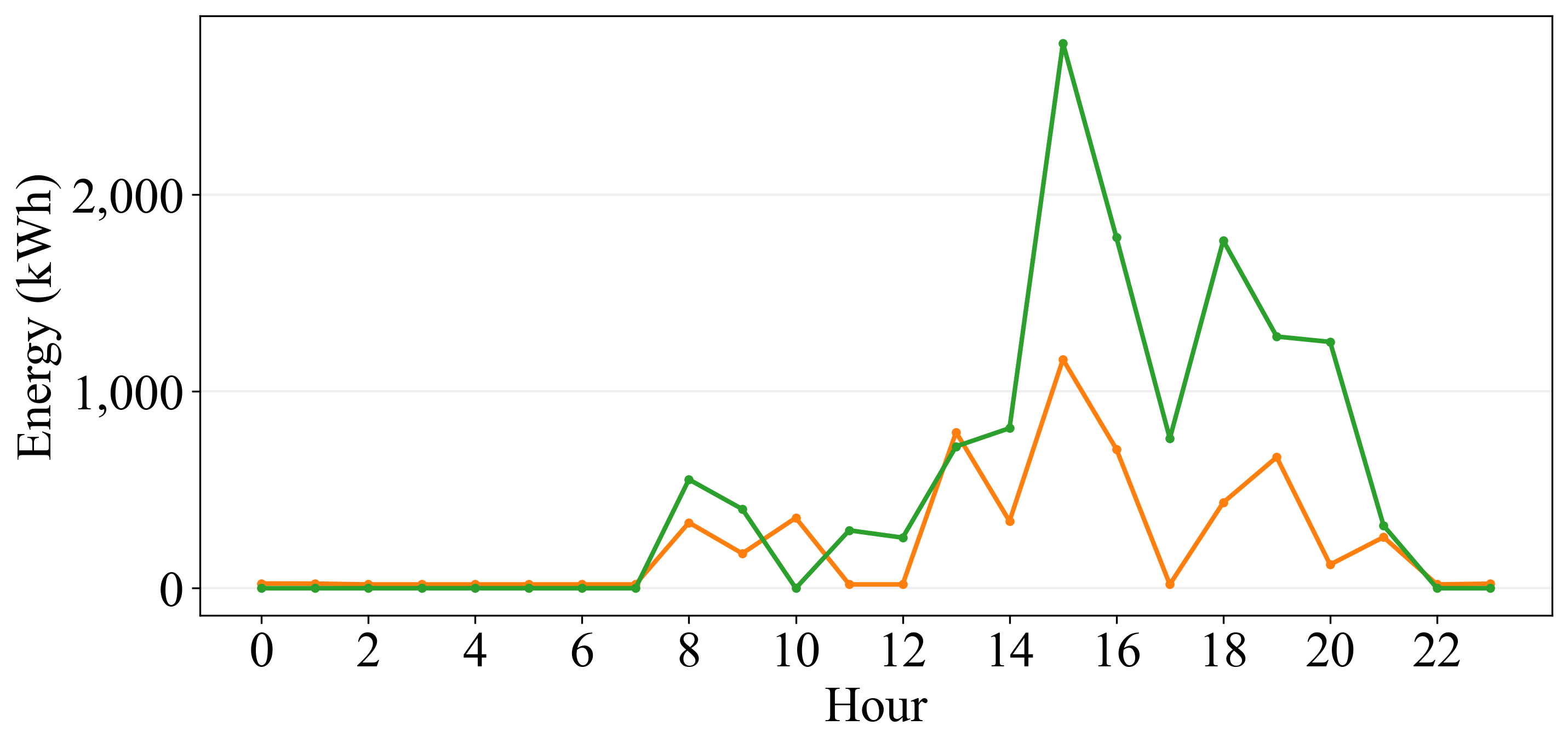}\caption{TNC Public -- Energy}\end{subfigure}\\[0ex]
  \includegraphics[width=0.8\linewidth]{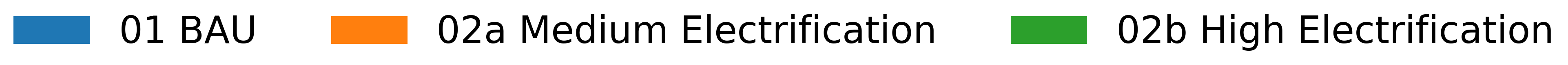}
  \caption{Hourly charging power (left) and energy (right) demand profiles by vehicle type and charging location across the three electrification scenarios.}
  \label{fig:charging_profiles}
\end{figure}

\Cref{fig:charging_profiles} presents hourly power and energy demand profiles across vehicle types and charging locations. Aggregate system energy peaks in the evening hours (7--9~pm) in the high electrification scenario at approximately 6~GWh, with corresponding peak power demand of approximately 4~GW. Personal home charging builds to an evening peak around 8--9~pm, reflecting post-commute charging behavior. Personal public charging exhibits a broader afternoon peak centered near 5~pm. Freight depot charging rises steadily from morning through evening, while dock charging peaks sharply around 1--3~pm, aligned with midday freight delivery schedules. TNC public charging volumes remain negligible across all scenarios.

\section{Conclusions}\label{sec:conclusion}

This paper presented a comprehensive 2050 scenario analysis for the Chicago metropolitan region using the POLARIS agent-based modeling framework, spanning vehicle electrification, freight operations, road pricing, and transit expansion. All policy scenarios were evaluated against a rigorously defined BAU baseline, enabling clean attribution of impacts to specific interventions.

The findings carry significant policy and planning implications for regional MPOs, utility providers, and municipalities. First, the spatial concentration of charging demand in the urban core, which drives a potential 4~GW evening peak under high electrification, underscores an urgent need for cross-sector infrastructure planning. Transportation agencies and electric utilities could consider co-optimizing grid upgrades and charging station deployments to prevent localized grid stress while supporting aggressive EV adoption targets. Second, the freight bundle results highlight that while operational mandates such as off-hour deliveries and rail diversion successfully reduce overall freight VMT, they can inadvertently fragment supply chains into more frequent, localized truck trips. Planners must pair delivery restrictions with urban consolidation strategies to prevent secondary congestion impacts. Third, as fleet electrification inevitably erodes traditional motor fuel tax revenues, our smart pricing results demonstrate that a spatiotemporally variable road user charge is a highly effective dual-purpose policy. It not only provides a structural replacement for infrastructure funding but also acts as the strongest lever to actively manage travel demand and reduce auto-dependence. Finally, to absorb the modal shifts prompted by such pricing mechanisms, ambitious capital transit investments—particularly the modeled rail extensions and BRT corridors—are critical to providing accessible, high-capacity mobility alternatives.

Several limitations should be noted. The scenarios are evaluated independently rather than in combination, limiting the ability to capture cross-policy synergies and conflicts. Future work should explore bundled scenarios that combine electrification, pricing, and transit interventions to identify complementary and competing effects. Additionally, extending the freight modeling to include e-commerce growth trajectories and last-mile delivery innovations would improve the fidelity of the freight scenarios. Finally, including land use changes considering transit-oriented development and evaluating its impact on transit ridership would be beneficial.

\section*{Acknowledgments} 
During the preparation of this manuscript, the authors used Claude Opus 4.6 and Claude Sonnet 4.6 through ARGO AI platform at Argonne National Laboratory to improve language, spelling, readability, literature review, coding, and result analysis. After using these tools, the authors meticulously reviewed and edited the entire text, code, and references (including manuscripts and all online links) and take full responsibility for the content of the publication.
\section*{Author Contributions} The authors confirm contribution to the paper as follows: study conception and design: Verbas, O., Altman, J., Beck, N., Cokyasar, T., Auld, J.; data collection: Altman, J., Beck, N., Zill, J., Nair, G.; methodology: Verbas, O., Cokyasar, T.; software: Verbas, O., Cook, J., de Camargo, P.V., Uhm, H., de Souza, F.; validation: Verbas, O., Altman, J., Beck, N.; analysis and interpretation of results: Verbas, O., Cokyasar, T., Alam, M.R., Bhidya, H.; visualization: Bhidya, H.; draft manuscript preparation: Alam, M.R., Cokyasar, T. All authors reviewed the results and approved the final version of the manuscript.
\section*{DECLARATION OF CONFLICTING INTERESTS} The authors declared no potential conflicts of interest with respect to the research, authorship, and/or publication of this article.
\section*{FUNDING} This material is based on work supported by the U.S. Department of Energy's Energy to Communities (E2C) program, which provides communities with expertise and tools to make local energy systems more affordable, reliable, and secure through three unique program offerings: in-depth partnerships, peer-learning cohorts, and expert match. E2C is funded by the U.S. Department of Energy and managed by National Renewable Energy Laboratory (NREL) with support from Argonne National Laboratory (ANL), Lawrence Berkeley National Laboratory (LBNL), Oak Ridge National Laboratory (ORNL), and Pacific Northwest National Laboratory (PNNL).


\newpage
\bibliographystyle{elsarticle-harv} 
\bibliography{trb_references}

\end{document}